\documentclass{JHEP3}
\pdfoutput=1
\usepackage{epsfig}
\usepackage{amsmath}

\newcommand{\be}{\begin{equation}}
\newcommand{\ee}{\end{equation}}
\newcommand{\la}{\langle}
\newcommand{\ra}{\rangle}

\newcommand{\normA}{\frac{N}{\pi}}
\newcommand{\normB}{\frac{N^2}{\pi}}
\newcommand{\normC}{N/2\pi}

\newcommand{\NA}{\mathcal{N}_a}
\newcommand{\NB}{\mathcal{N}_b}
\newcommand{\NC}{\mathcal{N}_c}

\newcommand{\NAn}{\NA^n}
\newcommand{\NBn}{\NB^n}
\newcommand{\NCn}{\NC^n}

\renewcommand\epsilon\varepsilon

\DeclareMathOperator{\diag}{diag}
\DeclareMathOperator{\tr}{tr}
\DeclareMathOperator{\re}{Re}
\DeclareMathOperator{\im}{Im}
\DeclareMathOperator{\prob}{Prob}

\title{Possible large-\boldmath{$N$} transitions for complex Wilson
  loop matrices} 

\author{Robert Lohmayer,$^a$ Herbert Neuberger,$^b$ and Tilo
  Wettig$^a$ \\ 
  \llap{$^a$}Institute for Theoretical Physics, University of
  Regensburg, 93040 Regensburg, Germany\\ 
  \llap{$^b$}Department of Physics and Astronomy, Rutgers University, 
  Piscataway, NJ 08855, USA\\
  Email: \email{robert.lohmayer@physik.uni-regensburg.de},
  \email{neuberg@physics.rutgers.edu},
  \email{tilo.wettig@physik.uni-regensburg.de}}

\abstract {It is shown that a very simple 
multiplicative random complex matrix model 
generalizes the large-$N$ phase structure 
found in the unitary case: A perturbative 
regime is joined to a non-perturbative regime at a point where
the smoothness of some quantities breaks down. A generic complex
Wilson loop matrix in a field theory 
admitting a 't Hooft planar limit could display a phase
transition in that limit as nonlinear effects become dominating over linear ones.}


\begin{document}

\cleardoublepage

\section{Introduction}

\subsection{Generalities}

Recent numerical work provides evidence
that Wilson loops in $SU(N)$ gauge theory in two, three and four dimensions
exhibit an infinite-$N$ phase transition as they are dilated from a 
small size to a large one. In the course of this dilation the eigenvalue distribution of the untraced Wilson loop unitary matrix expands from a small 
arc on the unit circle to encompassing the entire unit circle~\cite{ourjhep, three-d}.  The universality class of this transition is 
that of a random multiplicative ensemble of unitary matrices.
The transition was discovered by Durhuus and Olesen~\cite{duol} (DO) when they solved the
Migdal-Makeenko~\cite{makeenko} loop equations in two dimensional planar QCD. 
The associated multiplicative random matrix ensemble~\cite{janik} can be axiomatized in the
language of noncommutative probability~\cite{voicu}. It provides
a generalization of the familiar law of large numbers.
The essential feature making a difference is that 
one case is commutative and the other is not. 
Various recent insights into the DO transition
\cite{olesena, olesenb, blaizot} indicate 
that an even deeper understanding of the transition
might emerge. 
The large-$N$ transition does not imply confinement,
but if confinement occurs the large-$N$ transition is unavoidable if parallel transport
round small loops is close to the identity.

In this paper we relax the unitarity constraint.
We shall focus on a simple multiplicative random complex matrix model 
introduced in~\cite{cmplxrmt}, where it was shown that the model has an infinite-$N$ phase transition. The motivating physics application for
this study is to a more general gauge theory, obeying extra symmetries, 
which make complex matrix valued Wilson loop 
operators natural observables. If the
situation outlined above 
for ordinary gauge theories, where the matrix of the 
Wilson loop operator is unitary, generalizes to the complex case, 
the multiplicative random complex matrix model might capture the
universal features of large-$N$ transitions occurring in these 
more elaborate models. 

We also wish to point out that 
complex matrix transitions may also be relevant to ordinary gauge theories, in dimensions 
higher than two, for technical reasons. Ultraviolet divergences
of the bare Wilson loop
matrix may be eliminated by a regularization prescription that
makes the Wilson loop operator non-unitary. For concreteness, let us assume we are in four dimensions, dealing with
planar QCD. One example of a regularization is to 
associate with a curve ${\cal C}$ that has a marked point $x$ on it
the operator
\be
W({\cal C},x)=P e^{\!\!\underset{\cal C}{\;\;\;\int_x^x}\;
[i A\cdot dy +  \Phi(y) |dy|]}\:.
\ee
The ordered line integral starts at $x$ and follows the oriented 
curve ${\cal C}$ until it gets back to $x$. 
We take the extra scalar field $\Phi=\Phi^\dagger$ 
transforming as an adjoint under the gauge group, with a mass much  heavier than the QCD scale $\Lambda_{\text{QCD}}$. 
$A=A^\dagger$ is the usual gauge field, and $y_\mu(s)$ 
describes the curve ${\cal C}$.  
By adjusting the normalization of $\Phi$, its contribution could 
be made to cancel out the linear perimeter divergence
associated with $W$ but otherwise have little impact on smooth loops 
larger than the 
QCD scale on account of its large mass. For this to work $\Phi$
must enter the exponent without a $\sqrt{-1}$ prefactor. 
If the mass $M$ of $\Phi$ is very large,
one could extract the pure QCD string 
tension from loops of area ${\cal A}$
with $ 1\ll \Lambda_{\rm QCD} \sqrt{\cal A} \ll M/\Lambda_{\rm QCD}$. 
The regularization would make $W$ a finite operator, but one cannot 
associated it with a unitary matrix, and its spectrum would be spread
somewhat in the complex plane, defining a finite {\it surface}
eigenvalue density rather than a finite {\it linear} eigenvalue density.

\subsection{Outline of the paper}

We start by presenting a set of symmetries we require the complex
Wilson loop matrices to obey. These symmetries are inherited from
the Euclidean Gauge Theory producing the Wilson loops. We argue 
that these symmetries, in conjunction with the assumption that the
Wilson loop $W$ 
has a perturbative weak-coupling regime and a non-perturbative
``disordered'' regime, point to a large-$N$ phase transition in
the spectrum of $W$. The support of the eigenvalues of $W$ 
undergoes a topological change at the transition point, and this
indicates that there is something universal about the transition.
We shall refer to this hypothetical universality as large-$N$
universality, to distinguish it from ordinary critical phenomena
universality, which governs the structure of the pertinent Euclidean Gauge Theory at any finite $N$ and, by assumption, extends smoothly 
to infinite $N$. 

Armed with the universality assumption we then 
make a guess for the simplest
possible random matrix model which would be in the same large-$N$ universality class as the above Wilson loops. We proceed by 
discussing the general properties of the model. Some time 
is devoted to a technical point: Simplifications occur when one
drops the $\det W=1$ constraint, but as in usual planar QCD, dropping
the constraint has no impact on the infinite-$N$ phase structure.
In the context of the model we can be more specific, without actually
solving the model, about the shape of the support 
of the spectrum of $W$ at infinite $N$. We find that it is constrained
to an annulus in the complex plane whose internal and external radius
are reciprocal. As the coupling or loop size of $W$ change the
spectrum evolves from a simply connected small blob centered at
$z=1$ to a multiply connected region. The annulus the spectrum is
confined to also expands. This picture mirrors the picture we presented
in the Euclidean Gauge Theory case described earlier.

We next proceed to a more detailed analysis of the model using the
average of the modulus square of the characteristic polynomial of $W$.
This observable is shown quantitatively 
to produce a spectrum with the properties that
were anticipated. We use analytical and also some numerical tools.
The main analytical tool is a representation of the observable in terms
of an integral over Grassmann variables with a local action in
an internal space; the Grassmann variables are akin to mathematical quarks. After the introduction of the Grassmann variables the matrix
averaging of the model can be done explicitly and, eventually,
the entire dependence on $N$ becomes explicit. This sets the stage
for a saddle-point analysis at infinite $N$. We perform the analysis
only to the extent that it gives the phase structure. Global stability
questions are dealt with by numerical tests and not by purely
analytical methods. 

To get some feeling for the universality of the model we proceed with a
slight generalization. This generalization has an extra parameter which
allows an interpolation between the complex multiplicative matrix
model and the unitary multiplicative matrix model that has 
been extensively studied in past work. In that, the generalized model
provides further support to the view that in some sense the large-$N$
transition here has a direct relationship to the large-$N$ transition
found in ordinary gauge theories with unitary Wilson loop matrices. 
It is seen that, similarly to the original complex matrix model and
to the unitary matrix model, the inviscid Burgers equation plays a
central role also in the generalized model. 

The last part of the paper is only partially successful. Although the
infinite-$N$ phase structure justifies the guess that there is a
large-$N$ universality class associated with it, to make this concrete
one needs to go to sub-leading terms in the large-$N$ expansion and
identify the (hopefully few) relevant variables. We note that for 
finite $N$ the matrix models get mapped into matrix models consisting 
of products of $2\times 2$ matrices. More precisely, the average
of the modulus square of the characteristic polynomial of the $N\times
N$ complex matrix $W$ can be exactly represented by the solution of
an associated multiplicative matrix model where the matrices are only
$2\times 2$ and the dependence on $N$ is explicit. 
However, this still leaves too many variables (albeit a finite, $N$-independent, number), impeding
an explicit analysis of the approach to the large-$N$ limit. 
Suspecting that the number of variables can be further reduced 
by dropping sub-leading corrections in large $N$ we simplify the
$2\times 2$ model further focusing on some special cases. We finally
present a case where we end up with only two real variables and
show how that model could be exactly solved. However, the exact solution
is in the form of an infinite series, and the study of the large-$N$
limit still presents difficulties. 

We end up being forced to leave further work on the 
large-$N$ universality class to the future, but feel that we have made
substantial progress here to eventually achieve a complete
understanding of the large-$N$ universality class.

The paper ends with a brief summary.

\section{General properties of our random matrix products}

The complex Wilson loop matrix is denoted by $W$. We assume that the Euclidean Field Theory which
defines $W$ provides a probability distribution for $W$, $P(W)d^{2N^2-2} W$, with some natural
properties:
\begin{itemize}
\item $\det W=1$,
\item $P(W)=P(W^{-1})=P(W^*)$,
\item $P(W)=P(UWU^\dagger)~{\rm for}~U\in U(N)$.
\end{itemize}
A construction of $W$ in terms of traceless double indexed fields
and discrete symmetries like parity and charge conjugation can assure the first two properties.
Gauge invariance implies the third property. 

To study the spectral properties of $W$, we define
\be
Q(z,z^*)=\langle |\det(z-W)|^2 \rangle,
\ee
where $\langle \ldots \rangle$ denotes averaging with respect to $P$. 
The general properties imply 
\be
Q(z,z^*)= Q(z^*,z)=|z|^{2N} Q(1/z^*,1/z)=|z|^{2N} Q(1/z,1/z^\ast)\:.
\ee
$P$ and consequentially $Q$ are assumed to depend on a coupling $\lambda\ge 0$ ($\lambda$ can be a running coupling, depending on
the size of the loop). When $N\to\infty$ 
$\lambda$ will stay finite, scaling with $N$ to produce the 't Hooft 
topological classification of diagrams. We assume that we have employed 
a regularization that respects the above symmetries, admits a standard
large-$N$ expansion and is in general benign in the sense that 
a large-$N$ transition will survive the continuum limit. 

Generically, $N\times N$ random matrices have regions in the spectral 
plane where the eigenvalue density is exponentially suppressed as 
$N\to\infty$; hence, at $N=\infty$ 
the eigenvalue density vanishes in these
regions. The eigenvalue density at infinite $N$ cannot 
vanish everywhere, so the large-$N$ limit induces some lack of smoothness
in the eigenvalue density. Typically, the eigenvalue density is 
guaranteed to be non-zero somewhere in the plane (rather than
disappearing at infinity) because for a small loop, $W$
is close to the identity matrix. In many theories one can replace
the phrase ``small loop'' by ``small $\lambda$''. However,
the condition $\det W=1$ ensures that 0 is not an eigenvalue. 

We therefore assume that 
in the $N=\infty$ limit, for any
$\lambda\ne 0$ the probability of $W$ having eigenvalues within some small finite circle around $z=0$ is zero. The radius of the circle increases to unity when $\lambda\to 0$
when all the eigenvalues of $W$ are forced into a shrinking region  around $z=1$. Typically, this
is reflected in $Q(z,z^* )$ having a holomorphic factorized form for $|z|<\rho(\lambda) < 1$,
\be
Q(z,z^*)=|f(z)|^2\:.
\ee
By the inversion symmetry eigenvalues are also excluded from around complex infinity, so
the complex plane can be thought of as a two-dimensional sphere with the north and 
south poles excised. At infinite $N$ the eigenvalues make up a connected region 
containing $z=1$ for any $\lambda$.  For 
$\lambda \ll1$ this region is very small and does not wrap around the doubly punctured sphere.
 
When $\lambda\to\infty$, the dynamics of the particular model become
important. We are interested here in the case where strong coupling
induces strong disorder in the Wilson loop spectrum. This is the case
in confining theories, but the evolution to 
sufficiently strong disorder causing a large-$N$ 
transition is not a compelling reason for the onset of confinement in more general theories. Strong disorder 
would imply that as the coupling becomes stronger, 
$P(W)$ becomes less restrictive, and the set of eigenvalues
of $W$ spreads widely. In this case it makes sense to 
assume the spectrum to completely surround the origin $z=0$. 
As a result the simply connected domain where the eigenvalues reside
at small couplings
becomes topologically nontrivial on the doubly punctured 
sphere, becoming multiply connected as a result of the
punctures. In principle, more complicated topology changes could happen, but, under some conditions, extra restrictions similar to the result of~\cite{feinzee} might apply,
leaving us with only the simplest option just described.  Intuitively, this is the generic way in which eigenvalues would spread 
out as disorder increases in a model obeying the general symmetries 
described earlier.

As already mentioned, $z=1$ will typically be in the domain of eigenvalues. Thus, the unit circle intersects the
set of possible eigenvalues and we could look for a signal of the transition on $|z|=1$. This signal
would be the entrance of the point $z=-1$  into the domain of eigenvalues 
as $\lambda$ is increased from $0$ through the transition point. 
The points $z=\pm 1$ are special because they are fixed points of the inversion symmetry. $z=1$ is
in the domain of eigenvalues for any $\lambda$, and the transition occurs when $z=-1$ also
joins. This description makes the similarity with the unitary matrix case clear.

\section{General considerations about a basic random complex matrix model}\label{sec_BasicModel}

We now set up a simple random matrix model for the complex Wilson loop matrix.
Basically we replace the true $P(W)$ by a much simpler one. 
The model is almost identical to that of~\cite{cmplxrmt}; the minor difference
is irrelevant, but shall nevertheless be touched upon later. 

The integration measure over complex numbers $z=x+iy$ is defined as 
\be
d\mu(z)=\frac12d^2z=dx dy\:.
\ee

\subsection[$SL(N,\mathbb C)$ case]{\boldmath $SL(N,\mathbb C)$ case}

Consider the space of traceless $N\times N$ complex matrices $C$ and define a normalized
probability density over it,
\be\label{RESprobdens}
P(C)d\mu(C)=e^{-N\tr C^\dagger C} \pi\delta(\tr C)\prod_{1\le i,j\le N} 
\frac{N}{\pi}d\mu(C_{ij})\:,
\ee
where the complex delta function is defined as $\delta(z)=\delta(x)\delta(y)$.
For any complex matrices $A$ and $B$ we have
\be\label{RESexpAB}
\int P(C)d\mu(C) e^{\tr C^\dagger A + \tr B^\dagger C}= e^{\frac{1}{N}
  \tr B^\dagger A-\frac1{N^2} \tr A \tr B^\dagger}\:.
\ee

Define now a sequence of $n$ i.i.d. matrices $M_j$,
$j=1,\ldots,n$, 
\be
M_j=e^{\epsilon C_j}\:,
\label{versaMC}
\ee
where $C_j$ is distributed by $P(C_j)$ and $\epsilon >0$ is a small
number. The delta function in the probability density (\ref{RESprobdens}) ensures $\det(M_j)=1$ for all $j=1,\ldots,n$.
The distributions of the $C_j$'s are invariant individually under
$C_j\to C^*_j, -C_j$ and $C_j\to U_j^\dagger C_j U_j,~U_j\in U(N)$. 

Define
\be
\label{versa}
W_n = M_1 M_2\cdots M_n=\prod_{j=1}^n M_j\:.
\ee
$M_j,M^*_j,M^{-1}_j$ and $U^\dagger M_j U,~U\in U(N)$ are equally probable.
Also, any two permuted sequences of $M_j$'s are equally probable. The distribution of
$W_n$ has the properties listed in the previous section. 

We are interested in the limit $n\to\infty$, $\epsilon\to 0$ with
$t=\epsilon^2 n$ held fixed at a non-negative value.\footnote{$\epsilon$ could have been complex, but $C\to e^{i\Phi} C$ can be used to make $\epsilon > 0$.} In that limit
the product matrix $W$ will be a finite matrix, and we are
interested in properties of its distribution as a function of $t$.

\subsection[$GL(N,\mathbb C)$ case]{\boldmath $GL(N,\mathbb C)$ case}

The matrices $W$ were strictly restricted to have unit determinant. Imposing the linear restriction $\tr C =0$ 
forces $W_n\in SL(N,\mathbb{C})$ for any sequence $M_j$.
However, little is lost by relaxing the determinant restriction. As we shall see later, this has no effect on the boundary of the region of non-vanishing eigenvalue density in the infinite-$N$ limit.

Without the restriction on the determinant, we would define a probability density (instead of (\ref{RESprobdens})) by
\be\label{probdens}
P(C)d\mu(C)=e^{-N\tr C^\dagger C}\prod_{1\le i,j\le N} \frac N\pi
d\mu(C_{ij})\:,
\ee
which would then lead to
\be\label{expAB}
\int P(C)d\mu(C) e^{\tr C^\dagger A + \tr B^\dagger C}= e^{\frac{1}{N}
  \tr B^\dagger A}\:.
\ee  
In this case, we are left with only the first term in the exponent of (\ref{RESexpAB}).

\subsection[Fokker-Planck equation and why $\det W=1$ does not matter at
infinite $N$]{\boldmath Fokker-Planck equation and why $\det W=1$ does
  not matter at infinite $N$}
\label{sec_FokkerPlanckDet}

In this section we drop the restriction $\tr C=0$ and explain why this does not affect the $N=\infty$ limit. The joint probability distribution of the entries
of the matrix $W_n$ in the limit $t$ fixed, for finite $N$, is $P_N(W_{\alpha\beta}; t)d\mu(W)$, where the $W_{\alpha\beta}$ are $N^2$ complex numbers
and $d\mu(W)$ is some conveniently chosen measure which is independent of $t$. 

We have a Markov chain~\cite{markov1,markov2}, and $P_N$ will satisfy a partial differential equation of the form~\cite{beni}
\be
\frac{\partial P_N}{\partial t}=\Omega_N P_N\:,
\label{littlefp}
\ee
where $\Omega_N$ is linear partial differential operator of at most second order acting
on the $2N^2$ real variables defining the $N^2$ complex numbers $W_{\alpha\beta}$. 
$\Omega_N$ depends explicitly on $N$ but has no explicit $t$-dependence. 

$\Omega_N$ is determined by the terms of order $\epsilon$ and $\epsilon^2$ in $M_n$ in
the recursion
\be
W_n=W_{n-1} M_n\:.
\label{recur}
\ee
Higher-order terms do not matter. (\ref{littlefp}) is an equation of the Fokker-Planck (FP) type. 
$P_N$ is determined by the initial condition at $t=0$, which in our case is a 
$\delta$-function with respect to $d\mu(W)$ concentrated at $W$ given by the unit matrix.

The FP equation for $P_N$ is derived by first expressing the step-$n$ probability density $P_N^{(n)}(W)d\mu(W)$ in terms of
$P_N^{(n-1)} (W^\prime ) d\mu(W^\prime )$, where $\epsilon^2$ is kept fixed. The computation is 
based on the linear recursion~(\ref{recur}),
\be
W=W^\prime M\:,
\ee
where $M=e^{\epsilon C}$ with $C$ distributed according to $P(C)d\mu(C)$. One expresses $P_N^{(n)}(W)d\mu(W)$ 
in terms of $W^\prime$ going to order $\epsilon^2$. $M$ acts linearly on the rows of
$W^\prime$, and therefore the Jacobian is given by the product of the Jacobians per row, which is
$|\det M|^{2N}=e^{2\epsilon \re \tr C}$. This takes care of the change in the measure. 
The expansion in $W^\prime$
around $W$ produces first-order derivatives of $P_N^{(n-1)}$ at order $\epsilon$ and $\epsilon^2$
and second-order derivatives at order $\epsilon^2$. After averaging over the matrix $C$, 
the measure terms of order $\epsilon^2$ give terms proportional to $P_N^{(n-1)}$, while measure terms of
order $\epsilon$ can combine with terms of order $\epsilon$ from the expansion of $P_N^{(n-1)}$ giving
first-order derivative terms in $P_N^{(n-1)}$. All second-order derivative terms in $P_N^{(n-1)}$ come from the
expansion of $P_N^{(n-1)}$ and not from the measure. 

This discussion simplifies if one chooses a measure
term that is invariant under the recursion. This is possible in many cases when the evolution is
on a group manifold. In our case it is convenient to parametrize $W\in GL(N,\mathbb{C})$ as $W=w{\tilde W}$
with $w^N=\det(W)$ and ${\tilde W}\in SL(N,\mathbb{C})$. Correspondingly we factorize the measure,
$d\mu(W) = d\mu(w)d\mu({\tilde W})$, where $d\mu({\tilde W})$ is right-invariant on
$SL(N,\mathbb{C})$. $P_N^{(n)}(W)$ also factorizes in these variables. This follows by induction in $n$, since
if $P_N^{(n-1)}$ is factorized so is $P_N^{(n)}$, and the initial condition also factorizes. Hence
\be
P_N^{(n)}(W) = P_N^{(n)1}(w) P_N^{(n)2} ({\tilde W})\:.
\ee
$P_N^{(n)1}$ is very easy to compute, as it comes from an abelian ensemble. 
Because of the choice for the measure term and because of the invariance of $M$ under conjugation
by $SU(N)$ elements, the FP equation is invariant under $W\to W U$ with $U\in SU(N)$. 

One can now restore the relation between $n$ and $\epsilon$ and take
$\epsilon\to 0$, $n\to\infty$ at fixed $t$, leading to
\be
P_N^{(n)}(w,{\tilde W})\rightarrow p_N(w;t) P_N ({\tilde W}; t)\:.
\ee
If $W$ were
a product of $SU(N)$ matrices, because of the invariance under right multiplication
by elements of $SU(N)$, the form of the operator $\Omega_N$ would be uniquely fixed up
to an overall constant to being the Laplacian on the $SU(N)$ group manifold, and the FP
equation would become the heat-kernel equation on $SU(N)$, leaving only an overall scale to be
determined.

Parametrizing $\det W=e^{u+iv}$ with $u,v\in\mathbb R$, it turns out that both $u$ and $v$ are normally distributed and their distributions are $N$-independent. Therefore, replacing $W$ by $\tilde W$ in averages of characteristic polynomials (cf. Sec. \ref{sec_AvChPoly}) only produces an unimportant prefactor in the large-$N$ limit.  
As a result, we see that we can use $GL(N,\mathbb{C})$ instead of $SL(N,\mathbb{C})$ 
without affecting the large-$N$ limit.
At subleading order in $1/N$ there are differences, but they are easily determined from
$p_N (w;t)$. So, little is lost by working with $GL(N,\mathbb{C})$ instead of $SL(N,\mathbb{C})$.

\subsection[The limit $n\to\infty$, $\epsilon\to 0$, $t=n^2\epsilon$
fixed and the subsequent $N\to\infty$ limit]{\boldmath The limit
  $n\to\infty$, $\epsilon\to 0$, $t=n^2\epsilon$ fixed and the
  subsequent $N\to\infty$ limit}

\label{sec_gamma}

Our objective is to determine the region in the $z$-plane populated by
eigenvalues of $W_n$. As a first step, we would like to find some
bounds delimiting this region. The region will have a sharp boundary
after the $N\to\infty$ limit is taken. Even without the restriction $\det W=1$, at large $N$, 
exact inversion symmetry gets restored. We expect that after all the
limits are taken we shall have
\be
e^{-\gamma(t)} \le |z| \le e^{\gamma(t)},~~~~z\in{\rm spectrum}(W_n)\:.
\label{annulus}
\ee
The function $\gamma(t)$ is positive for all $t>0$. We expect
$\gamma(t)$ to be finite for all finite $t$ and to increase monotonically
with $t$, because of the associated increase in disorder. By definition, 
at any fixed $t>0$, $\gamma(t)$ is the smallest positive number for which~(\ref{annulus}) holds almost surely, i.e., with probability 1. 
The annulus keeps the spectrum of $W_n$ 
away from the origin and infinity for any finite $t$. 
The major structural change that can happen as $t$ is increased from 0 is that the 
spectrum wraps round the annulus. For small $t$ the spectrum is a small blob round $z=1$;
the inversion and reflection invariance give it a kidney shaped appearance. As $t$ increases 
the blob has a larger annulus available and expands into it, until eventually it reaches
around it, at some finite critical $t$.

To get some feeling for why there is a $\gamma(t)$ at all and how it behaves we start from some small $\epsilon$, large $n$ and large $N$,
without committing at the moment to any special relations between these
numbers, although we really are interested in the situation $n\sim1/\epsilon^2 \gg 1$ and, although $N \gg 1 $, we
want $n \gg N$.

\subsubsection[$t\to\infty$]{\boldmath$t\to\infty$}

We first fix some $N\gg 1$, take some fixed $\epsilon^2 \ll1$ and 
study what happens as $n\to \infty$. In terms of our true interest this
means we are are trying to understand the asymptotic behavior of
$\gamma(t)$ for $t=n\epsilon^2$ going to infinity. 

If we take $n\to\infty$ at fixed $\epsilon^2 >0$ the classical 
results of F\"urstenberg~\cite{multrmt3, multrmt4} and a theorem by 
Oseledec~\cite{osel} apply. Some  
standard texts on the topic of random multiplicative matrix ensembles are~\cite{multrmt1, multrmt2}.
Our discussion is generally based on~\cite{multrmt3,multrmt4,multrmt1,multrmt2} 
and specifically on~\cite{conew,cohen}, but we do not aim 
here for mathematical rigor.

We start by defining the norm of a vector $v\in \mathbb{C}^N$ by
\be
\|v\| =\sqrt{\sum_{\alpha=1}^N|v_\alpha|^2}
\ee
and the matrix norm by\footnote{In Dirac notation, $v W_n \to \langle v | W_n$.}
\be
\|W_n\|=\sup_{\|v\|=1} \|v W_n\|\:,
\ee
identifying it with the square root of the 
largest eigenvalue of $W_n W_n^\dagger$. The results about random matrix products mentioned above apply to $W_n$.
For a fixed $W_n$ we have 
\be
\|W_n\| \ge e^{\gamma(t)}\:.
\label{ineqgam}
\ee
The inequality may be sharp in 
the above equation because $z$, the eigenvalue of
$W_n$ that has maximum absolute value and defines $\gamma(t)$, 
could be associated to an eigenvector that is 
very different from the maximum eigenvector of $W_n^\dagger W_n$ for
all asymptotic times $t$. 
However, we expect that our case is generic enough for
a conjecture of~\cite{goldhirsch} to apply, which would allow
us to replace the inequality sign above by an asymptotic 
equality at infinite $t$. For a discussion, see~\cite{multrmt2}, page 21.  If this 
is assumed, (\ref{ineqgam}) can be replaced by an equality. 

This assumption is non-trivial: For example, in the Ginibre ensemble~\cite{ginibre},
the left-hand side of (\ref{ineqgam}) equals twice the right-hand
side. However, in this case $W_n$ is not given by a product, but
is just a complex matrix $C$ distributed according to 
$\exp[-N\tr(C^\dagger C)]$, and no noncommutative matrix products are
involved.\footnote{If we replace
the Gaussian distribution of $C$ by a distribution
where each element is real non-negative, uniformly drawn 
from the segment $[0,1]$, (\ref{ineqgam}) does become an equality --
in this case $C$ has non-negative entries 
and the equality is a consequence 
of a Perron-Frobenius theorem; see chapter 8 in~\cite{horn}.}
This is very different from $W_n$ for large $t$, but 
is intuitively close to the situation for small $t$, the case addressed
in the next subsection. There, our estimate for $\gamma(t)$ will be
direct, without involving the norm $\|W_n\|$.
In any case, that (\ref{ineqgam}) becomes an equality as $t\to\infty$ 
will be confirmed by
both the analytical and numerical results presented later in the paper.

To study $\|W_n\|$ we need to know what happens to $v W_n$ for an arbitrary $v$ with $\|v\|\ne 0$. We define $v_i$, $i=1,\ldots,n$ by
\be
v_{i} = v \prod_{j=1}^i M_j 
\ee
and $v_0=v$. 
We are only interested in the ray specified by $v$. Let 
\be
S_v\equiv \frac{\| v M_1 \|}{\|v\|}\:.
\label{Sv}
\ee
Because of the invariance under conjugation by $U(N)$ elements, the distribution of $S_v$ induced by that of $M_1$ is independent 
of $v$. We now write
\be
\log\|v_n\|-\log\|v\|= \sum_{m=1}^n \log\frac{\|v_m\|}{\|v_{m-1}\|}\:.
\label{trick}
\ee
This is a trick used in~\cite{conew}. 

The terms in the sum on the RHS of (\ref{trick}) are i.i.d. real numbers for
any fixed values of $n$, $\epsilon^2$, $N$ by the same argument as below Eq.~(\ref{Sv}). 
Therefore, we can calculate the probability distribution of the LHS by
calculating the characteristic function $F(k)$ (see Eq.~(\ref{charfunc}) below) of one of the terms on the RHS, taking
the power $n$, and taking the inverse Fourier transform of that.

We reproduce the equation describing the source of randomness, ignoring the zero trace condition,
\be
M=e^{\epsilon C},~~~~P(C) = {\cal N} e^{-N\tr C^\dagger C}\:.
\ee
The random variables on the RHS of (\ref{trick}) are denoted by $x$,
\be
x=\log\frac{\langle v | M M^\dagger |v \rangle}{\langle v|v \rangle}\:,
\ee
and the characteristic function of the identical distributions is
\be
F(k)=\langle e^{ikx}\rangle_{P(C)}\:.
\label{charfunc}
\ee
The random variable on the LHS of (\ref{trick}) is denoted by $y$,
\be
y=\log\frac{\langle v_n | v_n \rangle}{\langle v|v \rangle}\:.
\ee
Its probability density is
\be
P(y) = \int \frac{dk}{2\pi}e^{-iky} [F(k)]^n\:.
\ee

We now expand $x$ in $\epsilon$ to order $\epsilon^2$ 
in the calculation of $F(k)$ and assume that the expansion
in $\epsilon$ can be freely interchanged with various integrals. 
An expansion to order $\epsilon^2$ is assumed to be all that is needed,
since an alternative treatment of the ensemble, based on a 
Fokker-Planck equation, would also need only expansions to
order $\epsilon^2$. 

By $U(N)$ invariance we can rotate the vector $v$ to point in the 1-direction,
\be
x=\log\biggl(\sum_{j=1}^N  |M_{1j}|^2 \biggr )\:.
\ee
Thus, the vector $v$ has dropped out completely. We shall reuse the symbol $v$ below. We now introduce some extra notation,
\be
\re C_{11}=u,\quad C_{j1}=v_j,\quad C_{1j}=w_j\quad \text{for}~~j=2,\ldots,N\:.
\ee
$v$ and $w$ are $(N-1)$-dimensional complex column vectors. We have to order $\epsilon^2$
\be
x=2\epsilon u +\frac{\epsilon^2}{2} (v^Tw + v^\dagger w^* ) + \epsilon^2 w^\dagger w \:.
\ee
To calculate $F(k)$ it suffices to know the distribution of $u$, $v$, $w$,
\be
P(u,v,w) ={\cal N}^\prime e^{-N(u^2+v^\dagger v + w^\dagger w )}\:.
\ee
The integral giving $F(k)$ is Gaussian and can be easily done, resulting in
\be
F(k)=e^{-\frac{\epsilon^2 k^2 }{N}} \left [\frac{1}{1-i\frac{k\epsilon^2}{N} +\frac{1}{4N^2} k^2 \epsilon^4}\right ]^{N-1}\:.
\ee
To the level of accuracy in $\epsilon^2$ at which we are working, we can write
\be
F(k)=e^{-\frac{\epsilon^2 k^2}{N}} e^{i\frac{\epsilon^2 k (N-1)}{N}}\:.
\ee
The characteristic function of $y$ is
\be
\langle e^{iky}\rangle = e^{-\frac{n\epsilon^2 k^2}{N}+in\epsilon^2 k (1-\frac{1}{N} )}\:.
\ee
Defining 
\be
{\hat y}=\frac{y}{n},
\ee
the inverse Fourier transform gives 
\be
P({\hat y} )=\sqrt{\frac{Nn}{4\pi \epsilon^2} } e^{-\frac{Nn}{4\epsilon^2} [ {\hat y} - \epsilon^2 (1-1/N) ]^2 } \:.
\ee
The  F\"urstenberg theorems~\cite{multrmt3,multrmt4} now tell us that almost surely
\be
\lim_{n\to\infty} \frac{1}{n} \log \| W_n \| = 
\frac{\epsilon^2}{2}\left(1-\frac{1}{N}\right )\:.
\ee
So far, $\epsilon^2$ and $N$ have been kept fixed. We therefore
conclude that for large enough $n$, 
\be
\|W_n\| \sim e^{\frac{n}{2}\epsilon^2 (1-1/N)}\:.
\ee
We now simply replace $n\epsilon^2$ by the large number $t$ and take
$N\to\infty$, which is a relatively harmless limit. We conclude that
\be
\gamma(t)\sim \frac{t}{2}~~~~{\rm for}~~t\to\infty\:.
\ee

\subsubsection[$t\to 0$]{\boldmath$t\to 0$}

We now wish to take $\epsilon\to 0$, keeping $n$ and $N$ large but fixed.
In terms of $t$, this would correspond to the asymptotic regime $t\to 0$.
Since $\epsilon$ is very small, we can try to expand to just linear
order in $\epsilon$, keeping $n$ and $N$ finite albeit at large values.
To linear order in $\epsilon$, the noncommutative aspect 
of the product is lost, and we can write
\be
W_n=e^{\epsilon\sum_{j=1}^n C_j}\equiv e^{\epsilon \sqrt{n} {\hat C}}\:.
\ee
By an $O(n)$ rotation one can show that the matrix
${\hat C}=\frac{1}{\sqrt{n}}\sum_{j=1}^n C_j$ is 
Gaussianly distributed. The distribution
is fixed by its average and variance,
\be
\langle {\hat C_{ab}} \rangle =0,~~~~~\langle |{\hat C_{ab}}|^2 \rangle =\frac{1}{N}\:.
\ee
Here, $ 1\le a,b\le N$ are matrix indices. 
${\hat C}$ is distributed exactly like in the Ginibre ensemble~\cite{ginibre}. For $N\to \infty$ we have 
\be
{\rm spectrum}({\hat C})=\{z; |z|\le 1\}\:,
\ee
giving
\be
\max_w\{|w|; w \in {\rm spectrum}(W_n)\}=e^{\epsilon\sqrt{n}}=e^{\sqrt{t}}\:.
\ee
We are therefore led to
\be
\gamma(t)\sim \sqrt{t}~~~~{\rm for}~~t\to0\:.
\ee

\subsubsection{$\gamma(t)$ for all $t$}

Our subsequent work confirms the findings in~\cite{cmplxrmt} 
which, in turn, imply the existence of 
an annulus with a $\gamma(t)$ 
obeying our considerations. For $N=\infty$ we shall find that the inverse
function to $\gamma(t)$, which we call $T(\gamma)$ with $\gamma>0$, is given by
\be
T(\gamma)=2\gamma\tanh\frac{\gamma}{2},~~~~T(\gamma(t))=t\:.
\label{Tgamma}
\ee
The previously presented asymptotic results are recovered.
The two regimes, $t\to 0$ and $t\to\infty$, differed in 
the order in $\epsilon$ one goes to. With either choice,
one obtains a finite expression if $t$ is finite, so the
truncation of the expansion in $\epsilon$ is self-consistent. 
When the full limit $n\to\infty$, $\epsilon\to 0$ 
is studied at fixed arbitrary $t$, going to order $\epsilon^2$
should reproduce both asymptotic results in $t$, and we shall see
that this indeed happens.
Note that the crossover between the two asymptotic regimes
occurs roughly where $\sqrt{t}=t/2$, which means $t=4$.
It will turn out that as $t$ increases the spectrum encircles the origin first at a critical value of $t=4$. In some rough sense,
this is the point where the lack of 
commutativity among the factors in the product of
matrices becomes qualitatively important. It no longer is
appropriate to think about the product as that 
of several matrices, each close to unity -- a perturbative  
viewpoint. 
Some observables made out of the matrix product 
evolve in $t$ smoothly, but 
others will develop a singularity.

\section{Average of products of characteristic polynomials}\label{sec_AvChPoly}
 
For general $N$ it is difficult to derive a closed formula for the distribution of all the matrix entries of $W$. We are interested in just 
the spectral properties of
$W$. Even this is difficult to obtain for arbitrary finite $N$.  
Partial information about the distribution of eigenvalues 
can be obtained from the averages of characteristic polynomials related to
$W$. These polynomials are generating functionals for various moments of the
eigenvalue distribution. We shall denote the averages over the $C_j$
by $\la \ldots\ra$. 
The calculation of some simple characteristic polynomials is feasible. 

Obviously, $\la\det(z-W)\ra$ carries no information since expanding in $z$
we see that only traces of products of $M$ factors appear. The latter
can be expanded in $\epsilon$ and yield only factors of $C_j$, all of
which vanish due to the invariance of $P(C)$ under $C\to C e^{i\Phi}$.
Hence,
\be
\la\det(z-W_n)\ra=(z-1)^N\:.
\ee

The first non-trivial case is 
\be
Q(z,z^*)=\la |\det(z-W_n )|^2 \ra\:,
\ee
and from now on we focus on the calculation of the above in the limit.
(By ``the limit'' we mean the limit $n\to\infty$, $\epsilon\to 0$
with $t=\epsilon^2 n$ held fixed.)

If one applies large-$N$ factorization to $\la |\det(z-W_n )|^2 \ra$ (i.e., assuming that the average of the product can be replaced by the product of the averages) one gets holomorphic factorization, and all eigenvalues seem to have to be unity. 
For any $t>0$, 
holomorphic factorization will hold for $z$-values close to 0 and 
$z$-values close to $\infty$. These two regions are outside the annulus
defined by $\gamma(t)$. 
The full holomorphic factorized regime penetrates the annulus and 
will be connected for $t$ small enough, but will split into two
disconnected components for $t$ larger than some critical value.
There are two disconnected components when the eigenvalue
support, always contained within the annulus 
defined by $\gamma(t)$, fully encircles the origin $z=0$.

\section{Saddle-point analysis of the basic random complex matrix model}
\label{sec_saddleptanalysis}

We wish to calculate $Q$ as a function of $t$ and see that at infinite $N$ 
the transition we are looking for indeed occurs. As a first step we need
some device to disentangle the nonabelian product defining $W$. Then, we can
make the $N$-dependence explicit by integrating out all degrees of freedom
whose multiplicity is $N$-dependent. This allows us to take $N\to\infty$
which, as usual, becomes a saddle-point problem. We analyze the saddle-point 
problem partially, only to the point where we can identify the transition we 
are after.

\subsection{Quark representation of characteristic 
polynomials for matrix products}

Let $X_1,X_2,\ldots,X_n$ be $n$ $N\times N$ square matrices of general structure.
We are interested in the characteristic polynomial of 
\be
W=X_1 X_2 \cdots X_n\:.
\ee

We introduce $nN$ pairs of Grassmann variables (quarks) :
$\{\bar\psi_j, \psi_j\}_{j=1,\ldots,n}$ with the convention
that $\psi_{n+1}\equiv \psi_1$, etc.
Let us define
\be
I_n(X_1,X_2,\ldots,X_n) = \int \prod_{j=1}^{n} [d\psi_j d\bar\psi_j ]
e^{w\sum_{j=1}^n \bar\psi_j\psi_j - \sum_{j=1}^n\bar\psi_j X_j \psi_{j+1}}\:.
\ee
We claim that
\be
I_n(X_1,X_2,\ldots.,X_n)= \det(w^n - W)\:.
\label{quark}
\ee
We prove this by induction in $n$. For $n=1$ the result is trivial.
Assuming $n\ge 2$ we integrate over the pair $\bar\psi_n \psi_n$ 
to derive a recursion relation,
\be
w^N I_{n-1}\left(X_1,X_2,\ldots.,X_{n-2},\frac{X_{n-1} X_n}{w}\right) = 
I_n (X_1,X_2,\ldots,X_n)\:.
\ee
It is now easy to check that if the claim holds for $n-1$ it holds also
for $n$, and this concludes the proof.

Obviously $\det(w^n - W)$ is invariant under cyclic permutations of the
matrices $X_j$. In the quark representation this invariance is proved
by ``translating'' the Grassmann pairs in their index $j$. 

\subsection[Making the dependence on $N$ explicit]{\boldmath Making
  the dependence on $N$ explicit}

We start by defining a complex number $\sigma$ which depends on $n$
such that
\be
\label{polar}
z=e^{n\sigma},~~~z=|z|e^{i\Psi},~~~\sigma=\frac{1}{n} \log |z| +i\frac{\Psi}{n}
\ee
with $-\pi\le\Psi < \pi$. 
We now introduce $4n$ Grassmann variables
$\bar\psi_j,\psi_j,\bar\chi_j,\chi_j$ and write
\be
|\det(z-W_n)|^2=\int\prod_{j=1}^n [d\bar\psi_j d\psi_j d\bar\chi_j
d\chi_j] 
e^{\sum_{j=1}^n (e^\sigma \bar\psi_j\psi_j + e^{\sigma^*}
  \bar\chi_j \chi_j) } e^{-\sum_{j=1}^n (\bar\psi_j M_j\psi_{j+1} + \bar\chi_j
M^*_j\chi_{j+1})}\:,
\ee
where $\psi_{n+1}=\psi_1$ and $\chi_{n+1}=\chi_1$. 
This can be rewritten as
\be
|\det(z-W_n)|^2=\int\prod_{j=1}^n [d\bar\psi_j d\psi_j d\bar\chi_j
d\chi_j] 
e^{\sum_{j=1}^n (e^\sigma \bar\psi_j\psi_j + e^{\sigma^*}
  \bar\chi_j \chi_j) } e^{-\sum_{j=1}^n (\bar\psi_j M_j\psi_{j+1} -\chi_{j+1}
M^\dagger_j\bar\chi_{j})}\:.
\ee
We now perform several integration variable changes. We first switch the
sign of $\chi_j$. Keeping the symbols $\bar\chi,\chi$ for the new variables we now
replace $\chi_j$ by $\bar\chi_{j-1}$ (with the understanding that $\chi_1\rightarrow
\bar\chi_n$) and also replace $\bar\chi_j$ by $\chi_{j-1}$ (again with
the understanding that $\bar\chi_1\rightarrow \chi_n$). When the integration 
measure for the new variables is written in canonical order, two $(-1)^N$ signs cancel out.
We are left with
\be
|\det(z-W_n)|^2=\int\prod_{j=1}^n [d\bar\psi_j d\psi_j d\bar\chi_j
d\chi_j] 
e^{\sum_{j=1}^n (e^\sigma \bar\psi_j\psi_j + e^{\sigma^*}
  \bar\chi_j \chi_j) } e^{-\sum_{j=1}^n (\bar\psi_j M_j\psi_{j+1} +
  \bar\chi_j M^\dagger_j\chi_{j-1})}\:.
\ee
The integral over the $C_1,\ldots,C_n$ now factorizes and can be done
explicitly to sufficient accuracy in $\epsilon$ to produce the
correct $t$-dependent limit. The following equalities ought to be understood in the sense that they hold up to terms which vanish as $n\to\infty$, $\epsilon\to0$ at $t=\epsilon^2n$ fixed.
As usually is the case in stochastic
calculations, we need to keep expressions correct to order
$\epsilon^2$, but not higher. We need to expand $M$ only to linear
order in $\epsilon$, as the next term in the expansion won't make a
contribution after the $C$ integration because of its phase invariance,
and we end up with
\be
\la e^{-\bar\psi M\psi^\prime-\bar\chi M^\dagger \chi^\prime }\ra =
e^{-\bar\psi\psi^\prime -\bar\chi\chi^\prime -\frac{\epsilon^2}{N}
  \bar\psi\chi^\prime\bar\chi\psi^\prime-\frac{\epsilon^2}{N^2}\bar\psi\psi^\prime\bar\chi\chi^\prime}\:.
\ee
We used the external source formula (\ref{RESexpAB}) and included an extra minus sign
obtained when a Grassmann variable was moved through an odd number of
other Grassmann variables. 

We now separate the quartic Grassmann terms into bilinears by
introducing scalar complex bosonic multipliers, $\zeta_j$ and $\lambda_j$, $j=1,\ldots,n$,
\begin{align}
e^{ -\frac{\epsilon^2}{N}
  \bar\psi\chi^\prime\bar\chi\psi^\prime}&=\NA \int d\mu(\zeta)
e^{-N|\zeta|^2} e^{-\epsilon ( \zeta \bar\psi \chi^\prime - \zeta^*
  \bar\chi \psi^\prime)}\:,\\
  e^{ -\frac{\epsilon^2}{N^2}
  \bar\psi\psi^\prime\bar\chi\chi^\prime}&=\NB \int d\mu(\lambda)
e^{-N^2|\lambda|^2} e^{-\epsilon ( \lambda \bar\psi \psi^\prime - \lambda^*
  \bar\chi \chi^\prime)}
\end{align}
with
\be
\NA=\normA\:,~~~~\NB=\normB\:.
\ee
The relative minus sign is needed to get the right sign in front of
the quadrilinear Grassmann term. The integration measure is $d\mu(\zeta )=d\re\zeta\: d\im\zeta$. 

The net effect was to replace the average over the complex matrix
$C$ by an average over the complex numbers $\zeta_j$ and $\lambda_j$,
with all the ``noise'' now originating from the variables
$\zeta_j$ and $\lambda_j$. This prepares the scene for making the dependence on $N$ explicit. 

The limit we are after can be obtained from 
\begin{align}
\la|\det(z-W_n)|^2\ra&=\NAn\NBn\int\prod_{j=1}^n [d\bar\psi_j d\psi_j d\bar\chi_j
d\chi_j d\mu(\zeta_j)d\mu(\lambda_j)] e^{-N\sum_{j=1}^n|\zeta_j|^2-N^2\sum_{j=1}^n|\lambda_j|^2}\nonumber\\
&\quad\times
e^{-\sum_{j=1}^n( \bar\psi_j \psi_{j+1}(1+\epsilon \lambda_j) 
+\bar\chi_j \chi_{j-1}(1-\epsilon \lambda_j^\ast)) }  
e^{\sum_{j=1}^n (e^\sigma \bar\psi_j\psi_{j} + e^{\sigma^*}
  \bar\chi_j \chi_{j}) } \nonumber\\
  &\quad\times 
e^{-\epsilon\sum_{j=1}^n (\zeta_j\bar\psi_j \chi_{j-1} - \zeta^*_j
\bar\chi_{j}\psi_{j+1})}\label{keya}\:.
\end{align}
One further change of Grassmann variables reduces the number of terms
that are not diagonal in the index $j$, $\psi_{j+1} =\psi_j^\prime$ and
$\chi_{j-1}=\chi_j^\prime$. Again, canonical ordering of integration measures leads to two $(-1)^N$ factors which cancel out. Dropping the primes on the new variables we
obtain
\begin{align}
\la|\det(z-W_n)|^2\ra&=\NAn\NBn\int\prod_{j=1}^n [d\bar\psi_j d\psi_j d\bar\chi_j
d\chi_j d\mu(\zeta_j)d\mu(\lambda_j)] e^{-N\sum_{j=1}^n|\zeta_j|^2-N^2\sum_{j=1}^n|\lambda_j|^2}
\nonumber\\ &\quad\times
 e^{-\sum_{j=1}^n( \bar\psi_j \psi_{j}(1+\epsilon \lambda_j) 
+\bar\chi_j \chi_{j}(1-\epsilon \lambda_j^\ast)) }  
e^{\sum_{j=1}^n (e^\sigma \bar\psi_j\psi_{j-1} + e^{\sigma^*}
  \bar\chi_j \chi_{j+1}) } 
\nonumber\\ & \quad\times
  e^{-\epsilon\sum_{j=1}^n (\zeta_j\bar\psi_j \chi_{j} - \zeta^*_j
\bar\chi_{j}\psi_{j})}\:.
\end{align}
Carrying out the integral over the Grassmann variables we get
\be
\la|\det(z-W_n)|^2\ra=\NAn\NBn\int\prod_{j=1}^n [d\mu(\zeta_j)d\mu(\lambda_j)] e^{-N\sum_{j=1}^n|\zeta_j|^2-N^2\sum_{j=1}^n|\lambda_j|^2}
{\det}^N\begin{pmatrix}A & B \\ C & D \end{pmatrix},
\ee
where
\be
A=e^\sigma T^\dagger -1-\epsilon \Lambda\:,~~ D= e^{\sigma^\ast}T-1+\epsilon \Lambda^\dagger\:,~~B=-\epsilon Z\:, ~~ C=\epsilon Z^\dagger
\ee
with
\be
T=\begin{pmatrix}0 &1 & 0 & \cdots & 0 & 0\\
0 &0 & 1  & \cdots & 0 & 0\\
\vdots & \vdots & \vdots &\cdots & \vdots & \vdots \\
0 &0 & 0  & \cdots & 0 & 1\\
1 & 0 & 0 & \cdots & 0 & 0\end{pmatrix}, ~~~Z=\diag(\zeta_1,\ldots,\zeta_n)\:, ~~~\Lambda=\diag(\lambda_1,\ldots,\lambda_n)\:.
\ee
Using known formulas on determinants of block matrices, we have
\be
\det\begin{pmatrix}A & B \\ C & D \end{pmatrix}=\det(AD-ACA^{-1}B)=\det A\det D 
\det ( 1+ \epsilon^2 Z^\dagger A^{-1} Z D^{-1} )\:.
\ee

\subsection[The trivial large-$N$ saddle and its domain of local
stability]{\boldmath The trivial large-$N$ saddle and its domain of
  local stability}

Since the $N$-dependence of the $\lambda$-integral is of the form
\be
\NBn\int\prod_{j=1}^n[d\mu(\lambda_j)]e^{-N^2\sum_{j=1}^n |\lambda_j|^2}{\det}^N\begin{pmatrix}A & B \\ C & D \end{pmatrix},
\label{lambdaInt}
\ee
we evidently get a trivial saddle point $\lambda_j=0$, $j=1,\ldots,n$
for large $N$ due to the dominance of the $N^2$ term. In this limit,
we can therefore focus on the remaining $\zeta$-integration with the replacements
\be
A\to A_0=e^\sigma T^\dagger-1\:, ~~D\to D_0=e^{\sigma^\ast}T-1=A_0^\dagger\:,
\ee 
which yields
\be
\la|\det(z-W_n)|^2\ra\to \NAn\int\prod_{j=1}^n [d\mu(\zeta_j)] e^{-N\sum_{j=1}^n|\zeta_j|^2}
{\det}^N\begin{pmatrix}A_0 & B \\ C & D_0 \end{pmatrix}.
\label{main_simple}
\ee
This expression is exactly equal to the result which we would obtain
without restricting the determinant of $W_n$. Integrals over complex
$\lambda$-variables were needed to decouple quartic Grassmann terms
arising from the second term in the exponent of (\ref{RESexpAB}),
which simply does not occur in (\ref{expAB}). The boundary of
non-vanishing eigenvalue density is therefore equal for $W_n\in
GL(N,\mathbb C)$ and $W_n\in SL(N,\mathbb C)$ in the large-$N$ limit.  

Using again the notations $A$ and $D$ for $A_0$ and $D_0$, we have
\be
\det\begin{pmatrix}A & B \\ C & D \end{pmatrix}=\det(AD-ACA^{-1}B)=|\det A|^2 
\det ( 1+ \epsilon^2 Z^\dagger A^{-1} Z (A^\dagger )^{-1} )\:.
\ee
The matrix $T$ implements cyclical one step shifts and 
obeys $T^n=1$. Hence we can write
\be
A^{-1}=\frac{1}{e^\sigma T^\dagger-1} =\frac{1}{1-e^{-n\sigma}} \sum_{s=1}^{n} e^{-s\sigma} T^s\:.
\ee
Each entry in $A^{-1}$ gets exactly one contribution from a single term in the
sum over $s$ above. $A^{-1}$ is a circulant matrix, which means that
it has identical entries on lines parallel to the main 
diagonal and periodic with period $n$ on lines parallel to the anti-diagonal.

The large-$N$ limit will obviously lead to saddle-point equations which
will be satisfied at $\zeta_j=0$ since the $\zeta_j,\zeta^*_j$ 
enter only bilinearly in the integrand.
If this saddle dominated at infinite $N$, we could replace 
$W_n$ by a unit matrix,
\be
\la |\det(z-W_n )|^2 \ra =|z-1|^{2N}\:.
\ee
This means there are no eigenvalues at any $z\ne 1$ in the complex plane. 
In particular, the eigenvalue density in the complex plane, scaled to be finite
at infinite $N$, is zero everywhere (except at the singularity $z=1$). We shall refer to
this saddle as saddle A. Where saddle A gives the correct answer,  
$\la |\det(z-W_n )|^2 \ra = |z-1|^{2N}$ is the
absolute value square of a holomorphic function in $z$ and there
is no finite eigenvalue surface density.  An eigenvalue surface density will develop 
in regions of the complex plane where saddle A is displaced by another saddle, saddle B, which destroys holomorphic factorization. 
To determine where saddle B must take over we find the regime where saddle A is no longer
locally stable. 
Comparison with numerical simulation shows that saddle A is always dominating whenever it is locally stable and that saddle B indeed
produces non-zero surface charge density. Thus, at the boundary of the domain of stability of saddle A
one has a continuous transition to regions with non-zero surface eigenvalue density. We do not calculate saddle B explicitly.

\subsubsection{Determination of the boundary of the domain of stability of the trivial saddle}

To determine the domain of local stability of the trivial saddle point we need
to calculate the Gaussian form of the integrand around saddle A. 
To quadratic order in $\zeta_j,\zeta^*_j$ we have
\be
\det\begin{pmatrix}A&B\\ C& D\end{pmatrix} = |\det(1-e^\sigma T^\dagger)|^2 
e^{\epsilon^2 F}\:.
\ee
We are interested only in $F$,
\be
F=\tr Z^\dagger A^{-1} Z (A^\dagger )^{-1}=\sum_{jl} \zeta^*_j 
|(A^{-1})_{jl}|^2 \zeta_l
\equiv \sum_{jl} \zeta^*_j K_{jl} \zeta_l\:.
\ee
Indices are understood modulo $n$, with the fundamental 
interval taken from $1$ to $n$. 
The matrix $K$ is also circulant, so its eigenvalues are determined by
the discrete Fourier transform of the last row of the matrix $K$ which
defines the entire matrix in terms of an $n$-term series,
$K_{nj}=K_j$, with
\be
K_j=\left |\frac{1}{1-e^{-n\sigma}}\right |^2 e^{-j(\sigma+\sigma^*)}\:.
\ee
The eigenvalues of $K$ are 
\be
\lambda_{k}=\left |\frac{1}{1-e^{-n\sigma}}\right |^2 
\sum_{j=1}^n e^{-j(\sigma+\sigma^*)} e^{\frac{2\pi i }{n} kj}=
\left |\frac{1}{1-e^{-n\sigma}}\right |^2  \frac{1-e^{-n(\sigma+\sigma^*)}}{1-
e^{\frac{2\pi i }{n} k} e^{-(\sigma+\sigma^*)}}
 e^{-(\sigma+\sigma^*)} e^{\frac{2\pi i k}{n}}\:.
\ee
Going back to the original variables, the condition for local stability is
\be
\epsilon^2 \frac{1}{|z-1|^2} \re \left ( \frac{|z|^2-1}{|z|^{\frac{2}{n}} e^{-\frac{2\pi i }{n}k} -1 } \right ) < 1
\ee 
for all $k=1,\ldots,n$. It is easy to see that if $|z| > 1$ the inequality is 
strongest for the $k=n$ case; the same is true for $|z|\le 1$. Hence, the 
condition holds also for all $k<n$ if it holds for $k=n$. We end up with a determination of the region of local stability of saddle A,
\be
\label{masterA}
\epsilon^2 \frac{1}{|z-1|^2} \left ( \frac{|z|^2-1}{|z|^{\frac{2}{n}} -1} \right ) < 1\:.
\ee 
Taking $n\to\infty$ at constant $t$ in Eq.~(\ref{masterA}) gives 
the chargeless region,
\be
1 > \frac{t}{2|z-1|^2} \frac{|z|^2-1}{\log |z|},
\label{regiona}
\ee
in agreement with Eq.~(83) of~\cite{cmplxrmt}. 

It is easy to see that the points on the boundary, separating charged and chargeless regions, having maximal or minimal absolute values are located on the positive real axis. This means that the function $\gamma(t)$, defined in Sec.~\ref{sec_gamma}, has to fulfill
\be
\gamma(t)=\frac{t}{2} \frac{e^{\gamma(t)}+1}{e^{\gamma(t)}-1}\:,
\ee
which is equivalent to Eq.~(\ref{Tgamma}).

Note that the exact 
invariance under inversion and complex conjugation of $z$ has been restored in the limit,
although it was lost at finite $n$ because of the truncation in the expansion in $\epsilon$
to second order (which was all that is needed to get the correct limit).  
Therefore, as explained earlier, one can look for the transition point by just
focusing on the unit circle. The portion of the unit circle which resides in the chargeless region
is
\be
t < |z-1|^2,~~~|z|=1\:.
\label{arcUnitCirc}
\ee
For $t<4$ there is an arc centered at $z=-1$ which resides in the chargeless region. The end points of this arc are at the
the two angles $\xi$ satisfying $\cos(\xi)=1-t/2$. 
When $t>4$ the charged region contains the unit circle and hence becomes multiply connected. 
The last point to be engulfed by the charged region as $t$ increases is the point $z=-1$.

\subsubsection{More detailed study of the neighborhood of the critical
point}

To better focus on the shape of the boundary on both sides of the
transition point $t=4$, it is useful to employ the following maps,
\be
z(u)=\frac{u+1/2}{u-1/2}\:,~~~u(z)=\frac{1}{2}\left ( {\frac{z+1}{z-1}}\right )\:.
\label{z-u}
\ee
$z=0$ and $z=\infty$ map into $u=\mp 1/2$, and $z=1$ maps into $u=\infty$.
There is always charge at $z=1$, so in the $u$-plane the charged region
extends to infinity. The $|z|=1$ circle maps into the imaginary axis in the $u$-plane, and $z=-1$ maps into the origin $u=0$. Inversion in $z$
becomes $u\to -u$. The real-$z$ axis maps into the real-$u$ axis.
The region $\{\im z > 0\}\cap \{|z|>1\}$ maps into the
region $\{\im u < 0\}\cap \{\re u > 0 \}$. 
Reflection about the real axis ($z\to z^*$) maps into reflection
about the real axis in the $u$-plane ($u\to u^*$), and reflection with
respect to the unit circle ($z\to1/z^*$) corresponds to
reflection about the imaginary axis in the $u$-plane ($u\to - u^*$). 
Our problem has these symmetries in the $z$-variable, so it suffices to
analyze one of the four quadrants in the $u$-plane and get the result for
the other quadrants by reflection through common axes in the $u$-plane. 

We have seen that eigenvalues are restricted to the annulus
\be
e^{-\gamma(t)} \le |z| \le e^{\gamma(t)}\:.
\ee
Therefore, the complement of this annulus is contained in the chargeless
region. Under the map, the two circles $|z|=e^{\pm \gamma(t)}$ go into two
circles with non-overlapping interiors in the $u$-plane,
\be
\left\vert u-\frac{1}{2}\coth\left(\pm\gamma(t)\right)\right\vert^2 = \frac{1}{(2\sinh\gamma(t))^2}\:.
\label{twocir}
\ee
The eigenvalues are restricted to the exterior of these two circles 
(the image of the annulus) and therefore the chargeless region~(\ref{regiona}) includes the interior of these two circles. 

Denoting $\re u = u_r$ and $\im u = u_i$, the chargeless region
is found to be
\be
0\le u_i^2 \le u_r \coth (t u_r) - u_r^2 - 1/4\:.
\label{bndyu}
\ee
For $u_r^2$ large enough the above inequalities self-contradict, showing
that the chargeless region is bounded in the $u$-plane. 

The chargeless region in the positive quadrant of the $u$-plane determines the entire chargeless region by reflections from quadrant to quadrant
through a common axis. When $u_r\to 0$ we have
\be
u_i^2 \le \frac{1}{t} - \frac{1}{4} + \left (\frac{t}{3}-1\right )
u_r^2 +O(u_r^4)\:.
\label{ur_0}
\ee
For $t<4$ there is a portion of the imaginary $u$-axis
inside the chargeless region; hence, the charged portion of the
imaginary $u$-axis has a break around the origin, and this maps
into the unit circle in the $z$-plane having a gap around $z=-1$. For $t<4$
there is one connected chargeless region containing the two circles
from Eq.~(\ref{twocir}). For $t>4$, $u_r=0$ is not in the chargeless
region. Consequently, the entire imaginary axis is in the charged region. 
The chargeless region is split into two disconnected regions,
each containing exactly one of the two circles of Eq.~(\ref{twocir}).

Exactly at the transition, when $t=4$, the boundary in the vicinity
of the origin is given by the two lines
\be
u_i=\pm \frac{1}{\sqrt{3}} u_r +\ldots
\ee
This critical contour is a slightly deformed figure-8, laying horizontally 
along the real-$u$ axis and symmetrically with respect to the imaginary-$u$ axis.
The midpoint of the 8, which resides at the origin of the $u$-plane, separates along the real $u$-axis as $t$ is decreased 
from $t=4$, splitting into two disconnected regions, 
and separates along the
imaginary-$u$ axis as $t$ is increased from $t=4$, becoming one single
connected region. 

\subsubsection{Connection to the inviscid Burgers equation}\label{sec_invBurgers}

The formula~(\ref{regiona}) for the boundary of the chargeless region
leads us to introduce the following map from the complex plane onto itself,
\be
Z(u,t)=\frac{u+\frac{1}{2}}{u-\frac{1}{2}} e^{-t u}\:.
\ee
For $\re u\neq 0$ Eq.~(\ref{regiona}) is equivalent to
\be
|Z(u,t)|
\begin{cases}>1 & \text{for}\ \re u(z) >0\:,\\ <1 & \text{for}\ \re u(z) <0\:,\end{cases}
\label{regionaa}
\ee
and the boundary between the chargeless and charged regions is given by
\be
|Z(u,t)|=1,~~\re u\neq 0\:.
\label{bndya}
\ee
For $\re u=0$, the boundary is found using (\ref{ur_0}). 

Only at $t=0$ is $Z(u,t)$ one-to-one. 
For nonzero $t$, $Z(u,t)$ has an essential singularity at $u=\infty$ 
which prevents an analytic definition of an inverse, $U(z,t)$. 
One is therefore led to look
for a local definition of the map, by a partial differential equation.
We differentiate the equation
\be
Z(U(w, t ), t )=w
\label{transca}
\ee
with respect to $w$ at fixed $ t$ and with respect to 
$ t$ at fixed $w$. We find then that $U(w, t)$ obeys 
\be
\frac{\partial U}{\partial  t} = U w \frac{\partial U}{\partial w}\:.
\ee
This is the inviscid complex Burgers equation (the Hopf
equation), up to a change of variables
$w=e^{-x}$. This equation plays a central role in two-dimensional YM and in the equivalent multiplicative random unitary
matrix model~\cite{duol}. It comes with the  
initial condition
\be
U(w,0)=\frac{1}{2}\left ( \frac{w+1}{w-1} \right )\equiv
u(w)=-\frac{1}{2} \coth \frac x2 \:.
\ee 
Defining $\xi(w, t)$ by
\be
\frac{U(w, t)+1/2}{U(w, t) -1/2}=e^{\xi(w, t)},~~~~U(w, t)=\frac{1}{2} 
\coth \frac{\xi(w, t )}{2}\:,
\ee
we find that~(\ref{transca}) is equivalent to
\be
e^{\xi(w, t)-\frac{ t}{2}\coth \frac{\xi(w, t)}{2}}=w\:.
\ee
This can be viewed as an equation for $\xi$, which in turn defines the
solution of the partial differential equation with the desired boundary condition.
The variables $x$ and $\xi$ are, up to some factors of $\pm i$, identical to the  variables denoted the same way in~\cite{duol}.

Equation~(\ref{bndya}) identifies the boundary separating the charged region
from the chargeless one with the image of the circle $|w|=1$ in the $u$-plane under the map $u=U(w,t)$ for $\re u\neq 0$. It is well known that as $ t$ increases from zero,
depending on the initial condition, singularities can be generated at finite $ t >0$. In our case, we have seen explicitly that at $ t=4$ a singularity
is generated.

\subsection{Precise relation to the model of
Gudowska-Nowak et al.}\label{sec_PreciseRel}

Instead of~(\ref{versaMC}), we could define
\be
\label{versb}
M_j=1+\epsilon C_j\:.
\ee
This is the exact form of the model of~\cite{cmplxrmt}. It looses the inversion symmetry
in $z$ at finite $n$. With this definition our formulas become exact even for finite $n$. 
The limit $n\to\infty$, $\epsilon\to\infty$
with $t=n\epsilon^2$ held fixed will not change if~(\ref{versa}) is replaced by~(\ref{versb}) and the inversion symmetry is recovered. 
However, with~(\ref{versb}) we can explicitly 
work out a few low-$n$ cases and test them either numerically or by more 
direct analytical means. We did this only for $n=1,2$. For these 
values of $n$ we obtained agreement with~\cite{cmplxrmt},
providing an additional check on our method, which relies only on information
captured by the observable $Q(z,z^*)$.

This ``linear'' model (as opposed to the previous, ``exponential'' 
version) is more convenient for numerical simulations, as one does not need
to exponentiate a large matrix.

\subsection{Numerical results}\label{sec_NumRes}

In general, we have not determined the global 
stability of our saddle, nor have we identified explicitly the competing nontrivial saddles. Therefore, we need a bit more evidence to establish the transition. We do this numerically.
Numerical checks were first carried out in~\cite{cmplxrmt}, and we confirm their results. 

We want to work out the inverse map $U(w, t )$ explicitly. First, we agree to
focus on one quadrant and use the symmetries to get the solutions in
the other quadrants.
We choose the quadrant $u_r, u_i > 0$, where $u_{r,i}$ are a short hand for the
real and imaginary parts of the function $U(w, t)$. 
Similarly, we denote the real and imaginary parts of
$v\equiv\frac12\frac{w+1}{w-1}$ by $v_r$ and $v_i$ respectively. We are given $v_{r,i}$ and need to find $u_{r,i}$.
The equation to be solved is
\be
\frac{u+1/2}{u-1/2} e^{- t u} =w=\frac{v+1/2}{v-1/2}\:.
\label{map_uw}
\ee
Here $ t$ is real but can have any sign, $u=u_r+iu_i$ and $v=v_r+iv_i$.
Taking the absolute value we get
\be
|u|^2+1/4=u_r \coth (b+ t u_r )\:,
\label{invabs}
\ee
where
\be
\tanh b = \frac{v_r}{|v|^2 +1/4}\:.
\ee
This gives $u_i$ in terms of $u_r$. The boundary of the chargeless region 
corresponds to $v_r=0$, see Eq.~(\ref{bndyu}).

A second relation is obtained by looking at the phase. We first write
\be
\frac{v+1/2}{v-1/2}=\left\vert\frac{v+1/2}{v-1/2}\right\vert e^{i\beta},~~~~
\frac{u+1/2}{u-1/2}=\left\vert\frac{u+1/2}{u-1/2}\right\vert e^{i\alpha}\:.
\ee
Then
\be
\alpha=\beta+ t u_i\:,
\ee
where, by convention, $|\alpha|,|\beta| < \pi$. $\beta$ is known, and we have
an expression for $u_i$ as a function of $u_r$ and other known quantities from
above. $\alpha$ is a function of $u_{r,i}$ so the above is a transcendental
equation for $u_r$ (or $\alpha$),
\begin{align}
\cos \alpha& = \frac{|u|^2-1/4}{|u^2-1/4|}\:,&\sin\alpha &=-\frac{u_i}{|u^2-1/4|}\:,&
\tan\alpha&=-\frac{u_i}{|u|^2-1/4}\:,\nonumber\\
\cos \beta &= \frac{|v|^2-1/4}{|v^2-1/4|}\:,& \sin\beta &=-\frac{v_i}{|v^2-1/4|}\:,&
\tan\beta&=-\frac{v_i}{|v|^2-1/4}\:.
\end{align}
The simplest form of the transcendental equation is obtained using the formulas
for the tangent, but the other equations are needed to resolve some discrete
ambiguities.

\begin{figure}
\centering
\includegraphics[height=0.49\textwidth,angle=90]{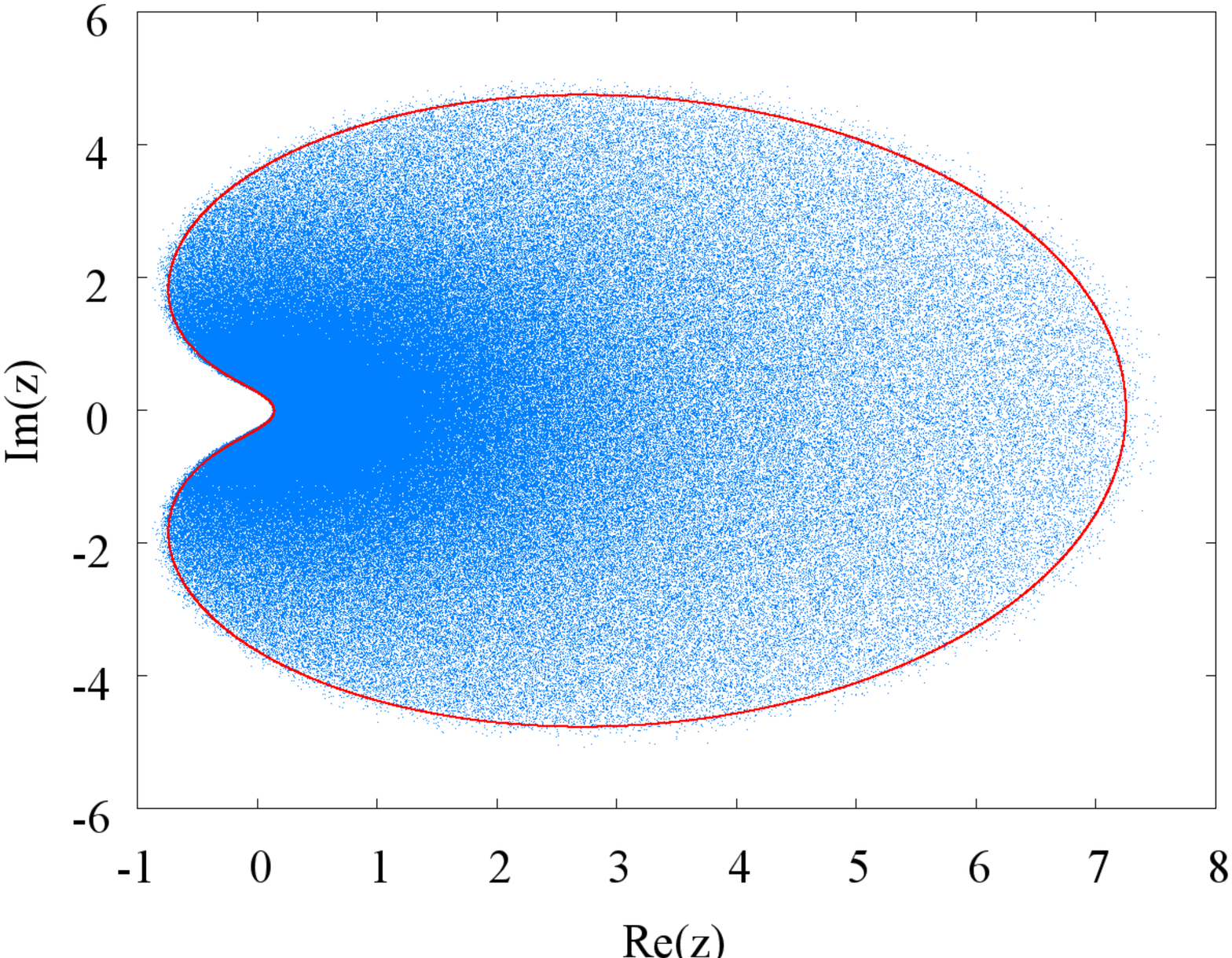}
\hfill    
\includegraphics[height=0.49\textwidth,angle=90]{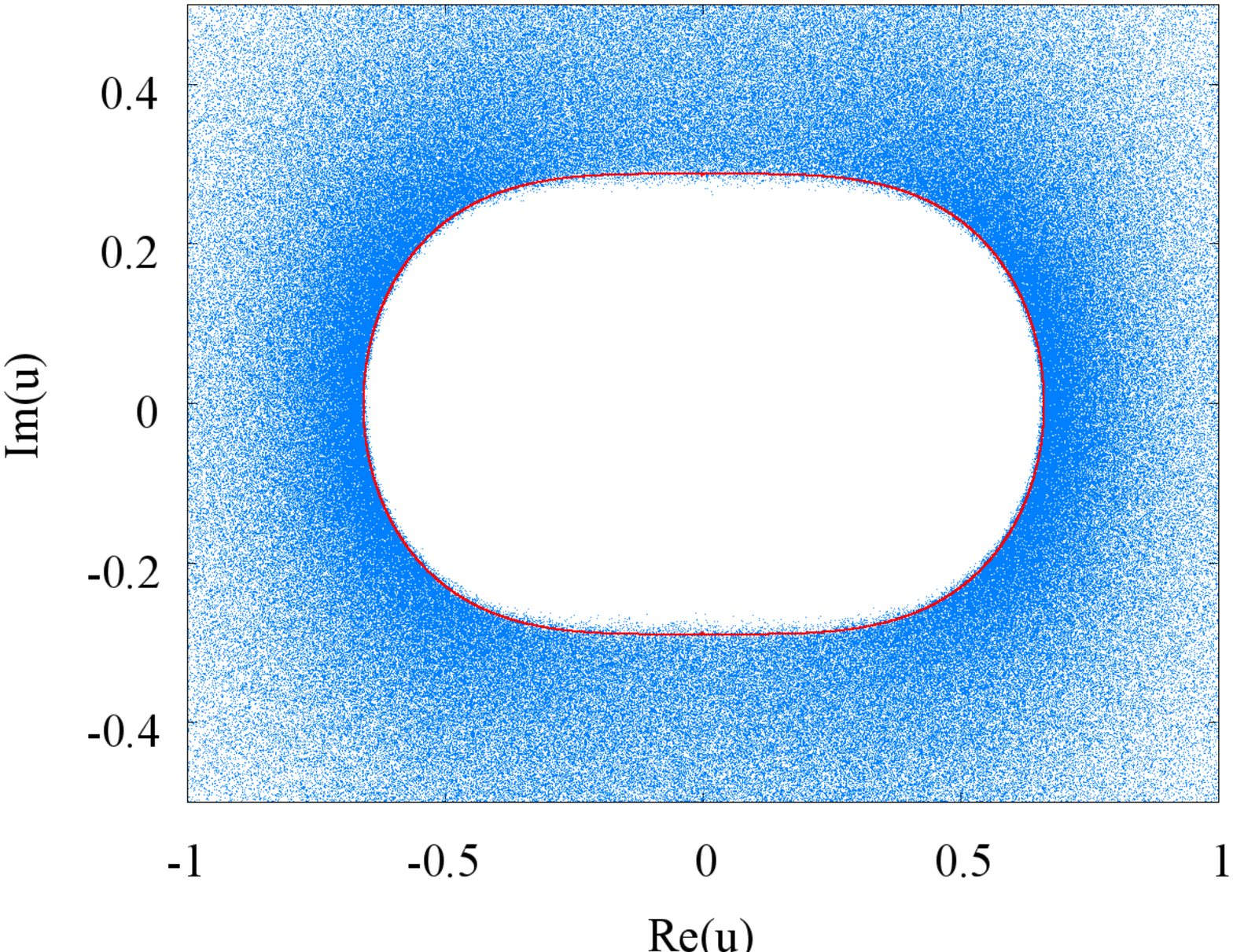} 
\\
\includegraphics[height=0.49\textwidth,angle=90]{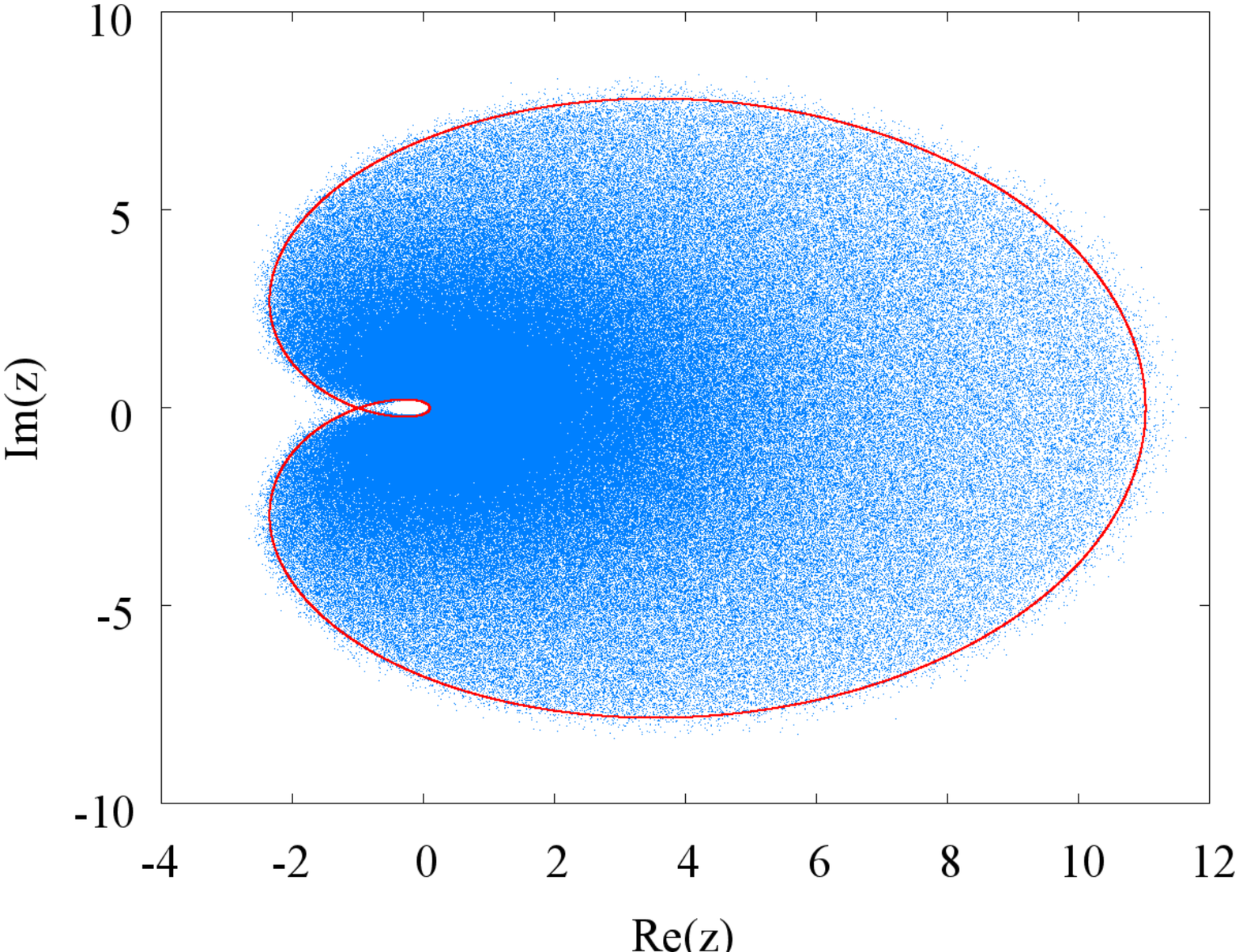}
\hfill    
\includegraphics[height=0.49\textwidth,angle=90]{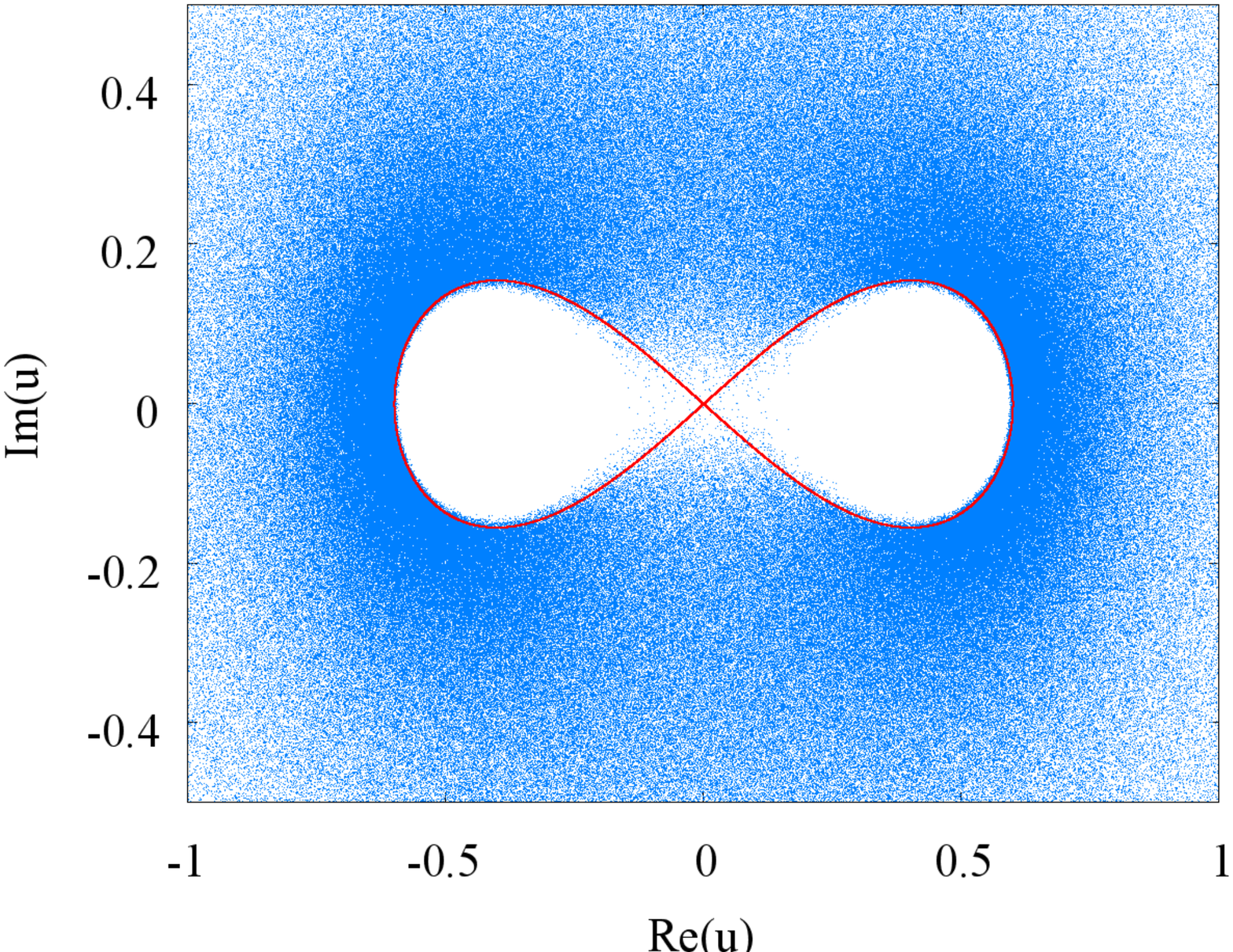} 
\\
\includegraphics[height=0.49\textwidth,angle=90]{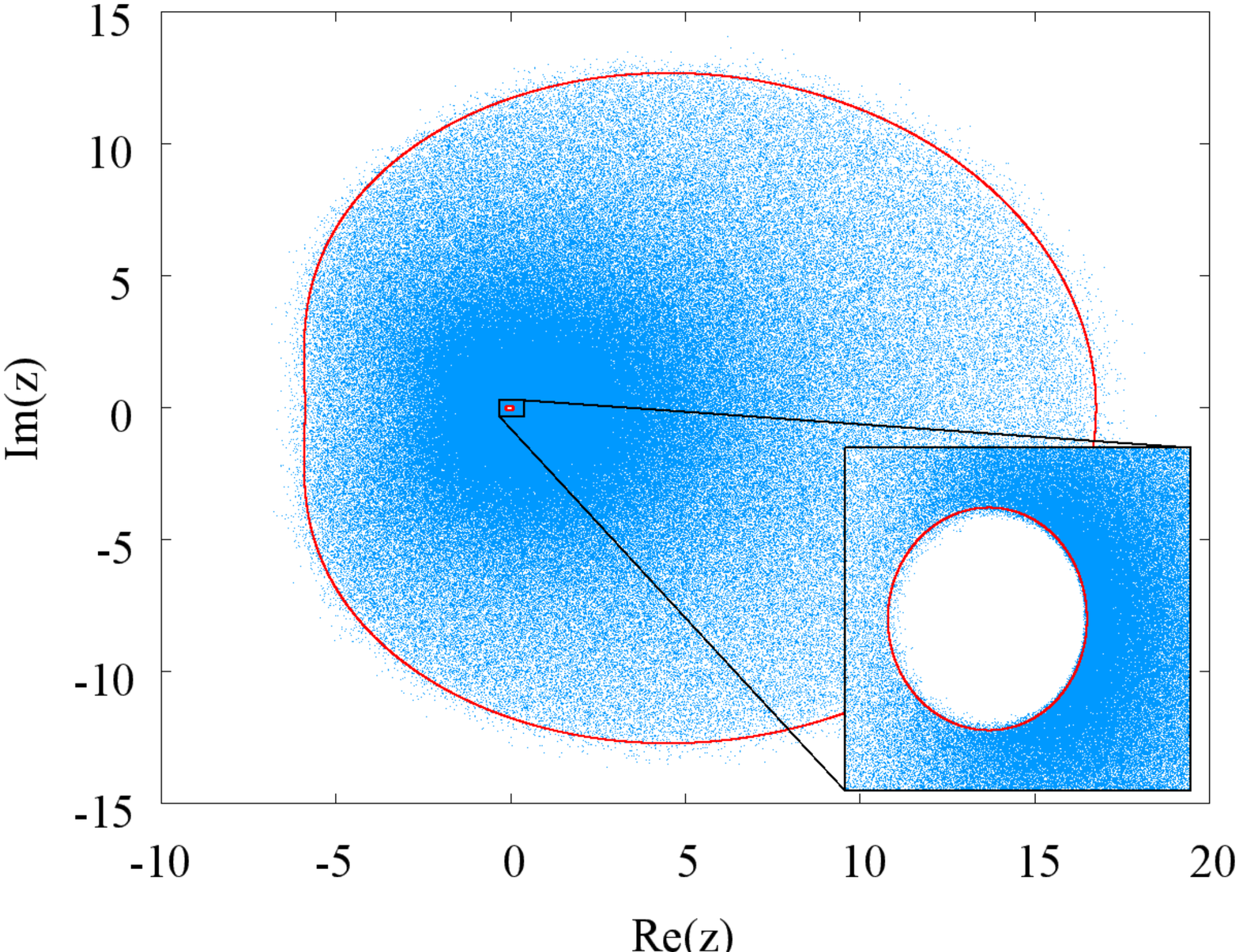}
\hfill    
\includegraphics[height=0.49\textwidth,angle=90]{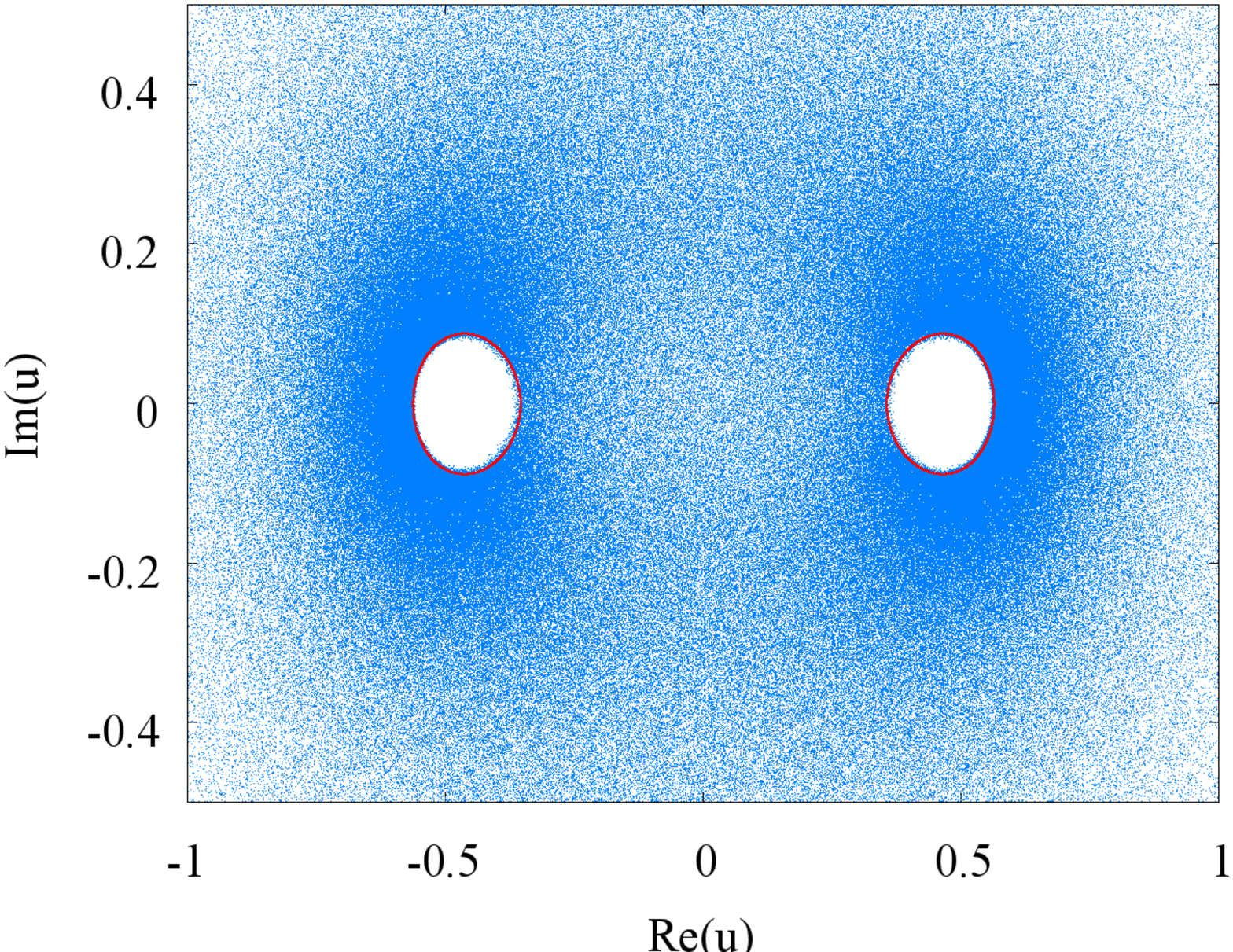}
 \caption{Scatter plot of the eigenvalues of $W$ in the $z$-plane (left) and in the $u$-plane (right) for  $t=3$ (top), $t=4$ (center), and $t=5$ (bottom).}
\label{fig1}
\end{figure}

All numerical simulations were performed with $n=2000$ and $N=2000$ for ensembles consisting of about 500 matrices (without a restriction on the determinant). In the figures below, the solid lines correspond to the analytically derived boundaries,  which are in very good agreement with numerical data. Figure~\ref{fig1} shows eigenvalue distributions in the $z$- and $u$-plane for $t=3$, $t=4$, and $t=5$. Numerical tests confirm that the topological transition of the domain of non-vanishing eigenvalue density occurs at $t=4$, when the domain becomes connected at $z=-1$. This corresponds to the imaginary $u$-axis completely lying in the domain of eigenvalues. Note also that the eigenvalue density in the $u$-plane is indeed symmetric under reflections at the real and imaginary axis, which is related to inversion symmetry in $z$.

Figure~\ref{fig2} shows eigenvalues for $t=12$ and affirms that for
large $t$ the boundary approximately consists of two circles with
center at $z=0$ and radii $\exp(\pm t/2)$.

\FIGURE{
\centering
\includegraphics[height=0.49\textwidth,angle=90]{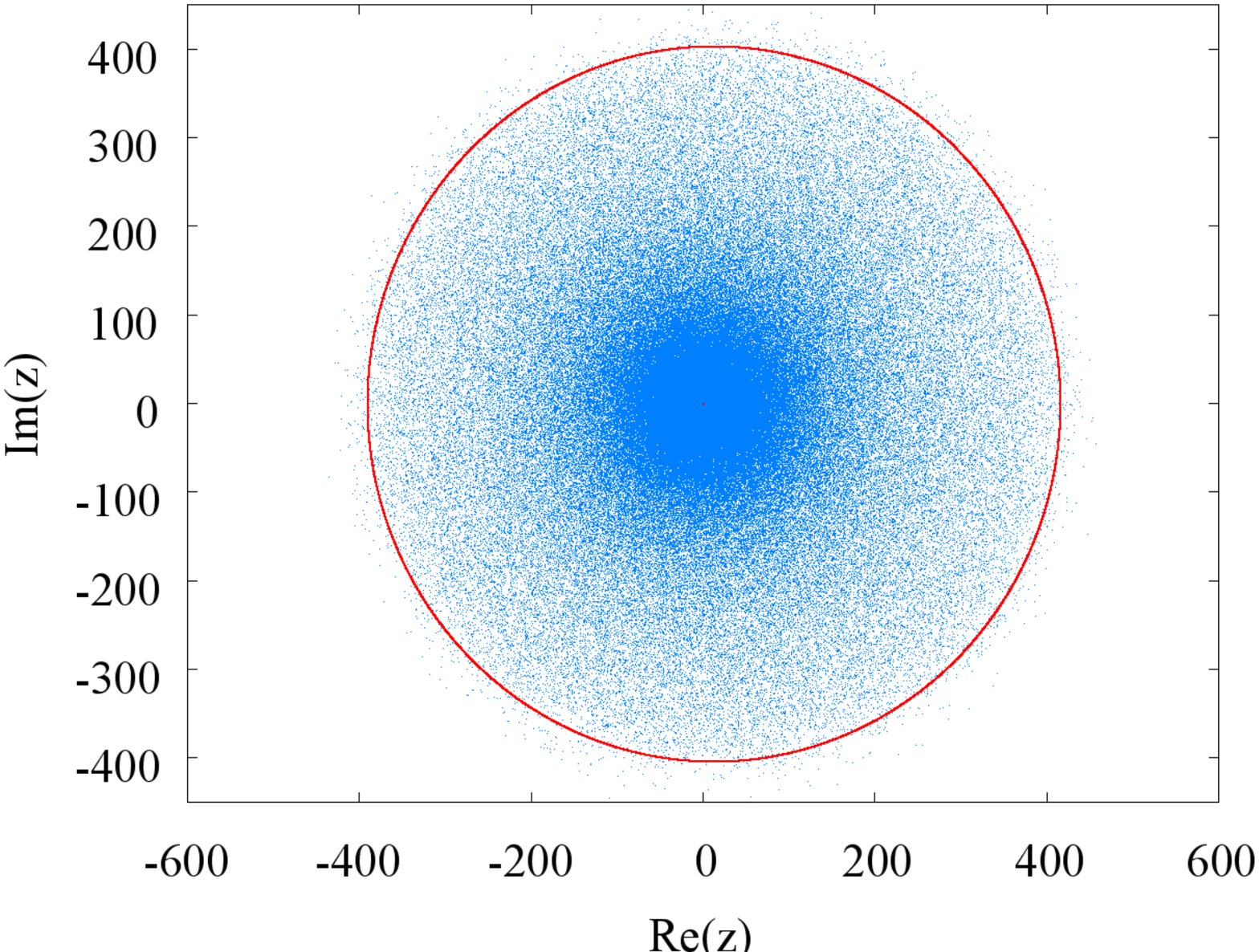}
\hfill    
\includegraphics[height=0.49\textwidth,angle=90]{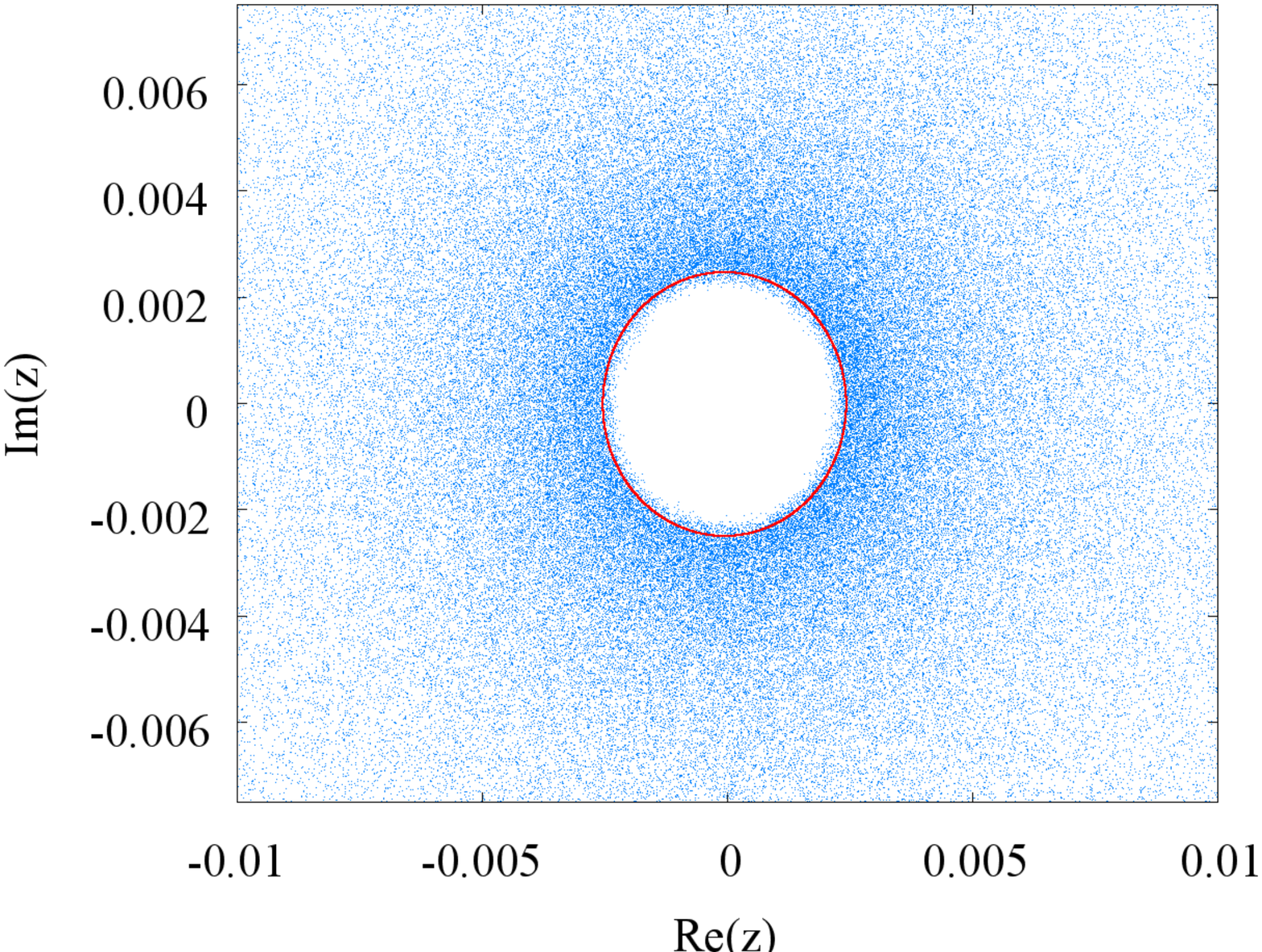}
 \caption{Plots of the eigenvalues of $W$ in the $z$-plane for
 $t=12$.  The plot on the right is an enlarged version of the plot on the left.}
 \label{fig2}
}

\section{A generalized Gaussian random complex matrix model}

\subsection{Definition and general properties}

The basic complex matrix model can be generalized to interpolate between
the case in which the individual factors are unitary and the case in
which they
are hermitian. Writing the matrix $C$ of each factor as $C=H_1+i H_2$
with $H_{1,2}$ hermitian and traceless, we take the probability distribution of $H_{1,2}$ to be
\be
P(H_1,H_2) = {\cal N}  e^{-N(\frac{1}{2\omega_1} \tr H_1^2 + \frac{1}{2\omega_2} \tr H_2^2)}
\ee
with $\omega_{1,2} > 0$. This produces the following correlation functions,
\be
\langle C_{ij} C_{lk} \rangle = \langle C^*_{ij} C^*_{lk} \rangle= \frac{1}{N} (\omega_1 - \omega_2 )\delta_{ik}\delta_{jl}-\frac1{N^2}(\omega_1-\omega_2)\delta_{ij}\delta_{lk}\label{correlCC}
\ee
and
\be
\langle C_{ij} C^\dagger_{lk} \rangle=\frac{1}{N} (\omega_1 + \omega_2 )\delta_{ik}\delta_{jl}-\frac1{N^2}(\omega_1+\omega_2)\delta_{ij}\delta_{lk}\:.
\ee
As a consequence, we have
\be
\Bigl\langle \bigl[ e^{\epsilon C} e^{\epsilon C^\dagger}\bigr]_{ij}\Bigr\rangle = \left(1+2\omega_1\left(1-\frac1{N^2}\right) \epsilon^2 +O(\epsilon^4)\right) \delta_{ij}\:.
\ee
Using the same analysis as in the original model we learn (see below) that the
spectrum of $W_n$, in the limit $n\to\infty$, $\epsilon\to 0$ with
$t=\epsilon^2 n$ held fixed, will now be restricted at any $N$ 
to the annulus\footnote{Now it would be meaningful to make $\epsilon$
complex, since for $\omega_1\ne\omega_2$ the phase invariance $C\to e^{i\Phi} C$ has been eliminated. One still has invariance under a sign
switch of $C$ and only $\epsilon^2$ enters. A complex $\epsilon$ would produce
a complex $t$.}
\be
e^{-\frac{2 \omega_1}{\omega_1+\omega_2}\gamma(t(\omega_1+\omega_2))} \le |z| \le e^{\frac{2 \omega_1}{\omega_1+\omega_2}\gamma(t(\omega_1+\omega_2))}\:,
\ee
where the function $\gamma$ is defined in Sec.~\ref{sec_gamma} and fulfills Eq.~(\ref{Tgamma}). 
As $\omega_1\to 0$ the spectrum is restricted to the unit circle 
$|z|=1$, and we are in the unitary case. For $\omega_1=\omega_2=1/2$
we are in the original model. For $\omega_2=0$ we are in the 
hermitian case, which produces a complex matrix upon multiplication of
the individual factors. 

\subsection[Large-$N$ factorized average]{\boldmath Large-$N$
  factorized average}

For $\omega_1\ne\omega_2$,
\be
\langle \det(z-W_n)\rangle = J(z)
\ee
is no longer equal to $(z-1)^N$, but to a more complicated
polynomial in $z$. The polynomial $J(z)$ 
is completely determined by the two-point function of the matrix $C$.

Where factorization holds we have
\be
\langle |\det(z-W_n )|^2 \rangle = |J(z)|^2
\ee
and no finite surface charge density. 

The polynomial $J(z)$ can be read off from previous work~\cite{three-d} on a product
of random unitary matrices. Take $\omega_1 < \omega_2$. The correlation functions in Eq.~(\ref{correlCC}) tell us that $J(z)$ depends
only on the difference $\omega_2-\omega_1$, so we could simply set
$\omega_1=0$, and then we obviously are in the unitary case 
with $C=iH_2$. So, for $\omega_2-\omega_1>0$ the dependence on
$\omega_2-\omega_1$ can be absorbed in $t$ by a rescaling,
\be
t_\pm\equiv t(\omega_2\pm\omega_1),~~~~t_+ \ge |t_- |\:.
\label{t_omega_diff}
\ee
From~\cite{three-d}, we get for the $SU(N)$ case
\be
J(z)=Q_N(z,t_- )=\sum_{k=0}^N 
\binom Nk z^{N-k} (-1)^k e^{-\frac{t^\prime  k(N-k)}{2N}}\:,
\ee
where $t^\prime = t_- (1+1/N)$. For large $N$ nothing is lost
by ignoring the difference between $t^\prime$ and $t_-$. From now
on we set $t^\prime=t_-$,
\be
J(z)=\sum_{k=0}^N 
\binom Nk z^{N-k} (-1)^k e^{-\frac{t_-  k(N-k)}{2N}}\:.
\ee
We also have an integral representation,
\be
J(z)= \sqrt{\frac{Nt_- }{2\pi}}
\int_{-\infty}^\infty d\lambda\:
e^{-\frac{N}{2}t_-  \lambda^2} 
\left[z-e^{-t_- (\lambda +1/2)}\right]^N \:.
\label{jza}
\ee
In~\cite{three-d} it was shown that for $t_- >0 $ the above polynomial
has all its roots on the unit circle. This was done by interpreting the integral
as the partition function of a classical ferromagnetic 
spin 1/2 model in an external magnetic field determined by $z$.   

For $\omega_2-\omega_1 <0$ we need to analytically continue to
negative $t_-$. As observed in~\cite{three-d} 
this is evidently possible in the polynomial
form. However, 
it no longer is true that all zeros are on the $|z|=1$ circle.
One can also analytically continue the integral expression,
\be
J(z)= \sqrt{-\frac{Nt_- }{2\pi}}
\int_{-\infty}^\infty d\lambda
e^{\frac{N}{2}t_-  \lambda^2} 
\left[z-e^{-t_- (i \lambda + 1/2)}\right]^N ,
\label{jzb}
\ee
where, by Eq.~(\ref{t_omega_diff}), now $t_-  < 0$.

Equations~(\ref{jza}), (\ref{jzb}) can be combined into one line-integral expression,
\be
J(z)= \sqrt{\frac{N t_- }{2\pi}}
\int_{\cal L} d\lambda\:
e^{-\frac{N}{2}t_-  \lambda^2} 
\left[z-e^{-t_- (\lambda +1/2)}\right]^N \:.
\label{jzc}
\ee
Here, $t_- $ is real of either sign and ${\cal L}$ is the real axis 
(from $-\infty$ to $\infty$) 
for $t_-  >0$ and the imaginary axis (from $-i\infty$ to $+i\infty$) 
for $t_-  <0$. For $t_- > 0$ we take $\sqrt{t_- } >0$, and for
$t_- <0$ we take $\sqrt{t_- }=-i\sqrt{-t_- }$ with $\sqrt{-t_- } >0$. 

\subsection{An exact representation of the generalized Gaussian model}

To determine for which values of $z$ the factorized formula no longer
holds and one expects non-zero surface charge density as a result of the
loss of holomorphic factorization we need the analogue 
of Eq.~(\ref{main_simple}). The method of getting at this formula is
the same; the one complication is that in addition to the complex noise
factors $\zeta_j$ and $\lambda_j$ one needs to introduce two additional real noise
factors $\xi_j$ and $\theta_j$. These extra noise factors are needed because
more quadrilinear Grassmann interaction terms need to be decoupled.
Still, at the end the dependence on $N$ is made explicit. 

It is also convenient to introduce the notation
\begin{align}
\omega_+&=\sqrt{\epsilon^2(\omega_1+\omega_2)}\:,\\
\omega_-&=\begin{cases}
\sqrt{|\epsilon^2(\omega_2-\omega_1)|} &\text{for}\ \omega_2>\omega_1\:,\\
i\sqrt{|\epsilon^2(\omega_2-\omega_1)|} &\text{for}\ \omega_2<\omega_1\:.\end{cases}
\end{align}
Since the integral over the variables $\lambda_j$ is again of the form (\ref{lambdaInt}) it can be trivially approximated by a saddle point at the origin in the large-$N$ limit. We are then left with
\be
\langle |\det(z-W_n)|^2 \rangle =\NAn\NCn\int\prod_1 ^n\left[ d\mu(\zeta_j)d\xi_jd\theta_j\right]e^{-N\sum_{j=1}^n\left(|\zeta_j|^2+\frac{1}{2}\xi_j^2+
\frac{1}{2}\theta_j^2\right)}{\det}^N
\begin{pmatrix} A&B\\ C&D \end{pmatrix},
\label{detABCD}
\ee
where $\NC=\normC$ and 
\begin{align}
A&= -\left(1-\frac12 \omega_-^2\left(1-\frac 1{N^2}\right)-\omega_-\sqrt{1+\frac{1}{N}}\Xi\right)+e^{\sigma}T^\dagger,\\
D&= -\left(1-\frac12 \omega_-^2\left(1-\frac 1{N^2}\right)-\omega_-\sqrt{1+\frac{1}{N}}\Theta\right)+e^{\sigma}T^\dagger,\\
 B&=-\omega_+ Z=-C^\dagger
\end{align}
with
\be
T=\begin{pmatrix}0 &1 & 0 & \cdots & 0 & 0\\
0 &0 & 1  & \cdots & 0 & 0\\
\vdots & \vdots & \vdots &\cdots & \vdots & \vdots \\
0 &0 & 0  & \cdots & 0 & 1\\
1 & 0 & 0 & \cdots & 0 & 0\end{pmatrix}
\ee
and 
\be
Z=\diag(\zeta_1,\ldots,\zeta_n)\:,\quad
\Xi =\diag(\xi_1,\ldots,\xi_n)\:,\quad
\Theta=\diag(\theta_1,\ldots,\theta_n)\:.
\ee

\subsection{Region of stability of the factorized saddle}

The identity
\be
\det\begin{pmatrix}A&B\\ C& D\end{pmatrix} = \det(A)\det(D) \det(1-D^{-1} C A^{-1} B)
\ee
shows that the variables $\zeta$ only enter as 
bilinears $\zeta_j\zeta_k^*$. At large $N$ the $\zeta$-integral will be
dominated by some saddle point, and $\zeta_j=0$ is a trivial solution to
the $\zeta$ saddle-point equations, for any $\Theta$, $\Xi$. 
Where this saddle point dominates, $z$ is in the chargeless region.
The reason is that we have $C=B=0$ at this saddle, and then the
remaining $\Xi$ and $\Theta$ integrals factorize and we have
\begin{align}
\langle |\det(z-W_n)|^2 \rangle& =\NCn \int\prod_{j=1}^n\left[ d\xi_jd\theta_j\right]e^{-N\sum_{j=1}^n\left(\frac{1}{2}\xi_j^2+
\frac{1}{2}\theta_j^2\right)}{\det}^N(AD)\nonumber\\
&=\Bigl\vert\NC^{\frac{n}{2}}
\int\prod_{j=1}^n\left[ d\xi_j\right]e^{-N\sum_{j=1}^n\left(
\frac{1}{2}\xi_j^2\right)}{\det}^N(A)\Bigr\vert^2\:.
\label{intDetA}
\end{align}
Since
\be
\det(A)=(-1)^{n-1}\left[z-\prod_{j=1}^n\left( 1-\frac{1}{2}\omega_-^2\left(1-\frac1{N^2}\right) -\omega_-\sqrt{1+\frac1N} \xi_j \right)\right ]
\label{detA}
\ee
depends only on $z$, and not on $z^\ast$, we have holomorphic factorization. Therefore, it is just the
structure of the $\zeta$-saddle which determines that $z$ 
is in a chargeless region. The holomorphic factor in Eq.~(\ref{intDetA}), however, is needed to
determine local stability. At the $\zeta=0$ saddle, from~\cite{three-d}
we know that in the $\epsilon\to 0$ limit we have
\be
\NC^{\frac{n}{2}}\int\prod_{j=1}^n\left[ d\xi_j\right]e^{-N\sum_{j=1}^n\left(
\frac{1}{2}\xi_j^2\right)}{\det}^N(A)=(-1)^{N(n-1)}J(z)\:.
\label{jzfull}
\ee
In the large-$N$ limit the simplified formulas~(\ref{jza}), (\ref{jzb})
apply. 

To determine local stability we need the saddle point which dominates
the integral over the $\xi_j$ in~(\ref{jzfull}). The derivation leading 
from (\ref{jzfull}) to (\ref{jza}) shows that for the case of
$\omega_2 > \omega_1$, at the saddle one has $\xi_j=\xi$, where
\be
\xi=\sqrt{\frac{t_- }{n}} \lambda_s
\ee
with finite $\lambda_s$. To leading order in $N$ we then have
\be
\det (A) \rightarrow (-1)^{n-1} \left [ z -\prod_{j=1}^n \left ( 1-
\frac{t_- }{2n} -\sqrt{\frac{t_- }{n}} \xi \right ) \right ] =
(-1)^{n-1} [z-e^{-t_-  (\lambda_s +1/2)} ]\:.
\ee
$\lambda_s$ is a saddle point of the integrand of~(\ref{jza}). With obvious
changes, a similar story holds for $\omega_2 < \omega_1$. 

We end up with the following expression for the matrices $A$ and
$D$ needed for the analysis of the stability of the trivial saddle 
under variations of $\zeta$,
\be
A=-\left[1-\frac{t_- }{n} (\lambda_s +1/2)\right] {\bf 1} + e^\sigma T^\dagger\:.
\ee
Here $\lambda_s$ has to satisfy 
\be
\lambda_s=\frac{1}{z e^{t_-  (\lambda_s + 1/2 )} -1}\:.
\ee
The appropriate contour ${\cal L}$ in Eq.~(\ref{jzc}), 
whose endpoints at infinity are fixed, will be deformed
to $\lambda_s$, and we assume that the integral will be dominated by one
single saddle point as long as one is in the chargeless region.  

It is now convenient to define ${\hat u}=\lambda_s+1/2$. Then the map
from $w$ to $z$ acquires a simpler form,
\be
z\equiv z(u)=\frac{{\hat u}+1/2}{{\hat u}-1/2} e^{-t_-  {\hat u}}\equiv Z({\hat u},t_- )\:.
\label{inverse}
\ee
In terms of the map $U(w, t)$ we have
\be
{\hat u}= U(z,t_- )\:,
\ee 
where we also allow for $t_- <0$, corresponding to ``backward evolution''. 

The stability of the trivial saddle is now determined by ${\hat u}$ from
\be
\det\left[1+\omega_+^2  Z^\dagger A^{-1} Z (A^\dagger)^{-1} \right],
\ee
with
\be
A=-e^{-\frac{t_-  {\hat u}}{n}}~{\bf 1}  + e^\sigma T^\dagger = e^{-\frac{t_-  {\hat u}}{n}}\left (-{\bf 1} +e^{\sigma+\frac{t_-  {\hat u}}{n}} T^\dagger \right )\:,
\ee
where, as before, $e^{n\sigma}= z$. We can drop the prefactor
$e^{-\frac{t_-  {\hat u}}{n}}$ because in the determinant there is
an extra $\epsilon^2$ prefactor. We therefore end up with
\be
\det\left[1 + \frac{t_+}{n} Z^\dagger {\hat A}^{-1} Z ({\hat A}^\dagger )^{-1} \right]
\ee
with
\be
{\hat A}=-{\bf 1} +e^{{\hat\sigma}} T^\dagger
\ee
and 
\be
{\hat z}\equiv e^{n{\hat\sigma}} \equiv z e^{t_- {\hat u}} = \frac{{\hat u}+1/2}{{\hat u}-1/2}\:.
\ee
Comparison with the $\omega_1=\omega_2=\frac{1}{2}$ case immediately 
leads to the region of stability (see Eq.~(\ref{regiona})),
\be
1 > \frac{t_+}{2|{\hat z}-1|^2} \frac{|{\hat z} |^2-1}{\log |{\hat z}|}\:.
\label{regionb}
\ee
${\hat z}$ is defined by the complex number ${\hat u}$ which 
should be determined by $z$ 
and $t_-$ in~(\ref{inverse}). Let the unit circle $|w|=1$ be parametrized by
$|s|\le \pi$, with $w=e^{is}$. For $|z|\neq1$, the boundary separating the charged and chargeless
region is defined in the $z$-plane by $z=f(s)$, given by
\be
f(s) = Z (U(e^{is},t_+ ) , t_- )~~~~\textnormal{for}~~\re U(e^{is},t_+ )\neq 0\:.
\label{fs}
\ee
Note that $t_+ \ge |t_-|$. For $t_+<4$, the boundary intersects the
unit circle in the $z$-plane at the points
\be
z=Z\left(\pm i\sqrt{\frac1{t_+}-\frac14},t_-\right).
\ee 
In complete analogy to the basic model, setting aside the restriction $\det(W_n)=1$ does not affect the boundary in the infinite-$N$ limit.

\subsection{Numerical results}

As mentioned above, the linear model is much more convenient for numerical simulations than the exponential one. Performing a similar stability analysis for the linear model, it turns out that the boundaries of the domains with non-vanishing eigenvalue density for the two models are equivalent up to a scaling by a factor of $\exp(-t_-/2)$.
The following figures show perfect agreement between numerically obtained eigenvalue domains and analytically determined boundaries for the linear model (data points as well as predicted boundaries are scaled by the corresponding factor of $\exp(-t_-/2)$). Therefore, we expect that the stability analysis gives the correct boundary for the exponential model, too.

\begin{figure}
\centering
\includegraphics[height=0.49\textwidth,angle=90]{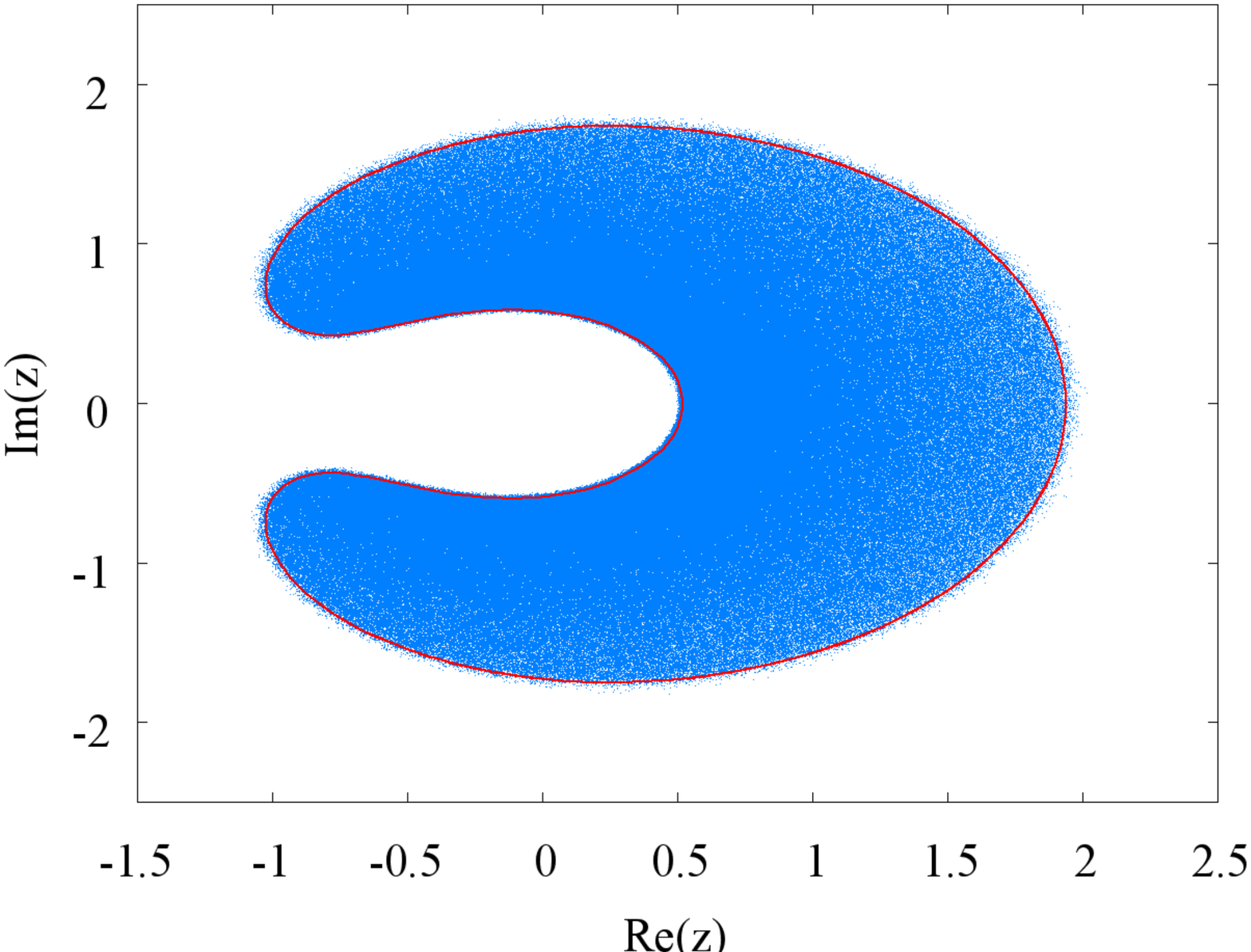}\hfill    \includegraphics[height=0.49\textwidth,angle=90]{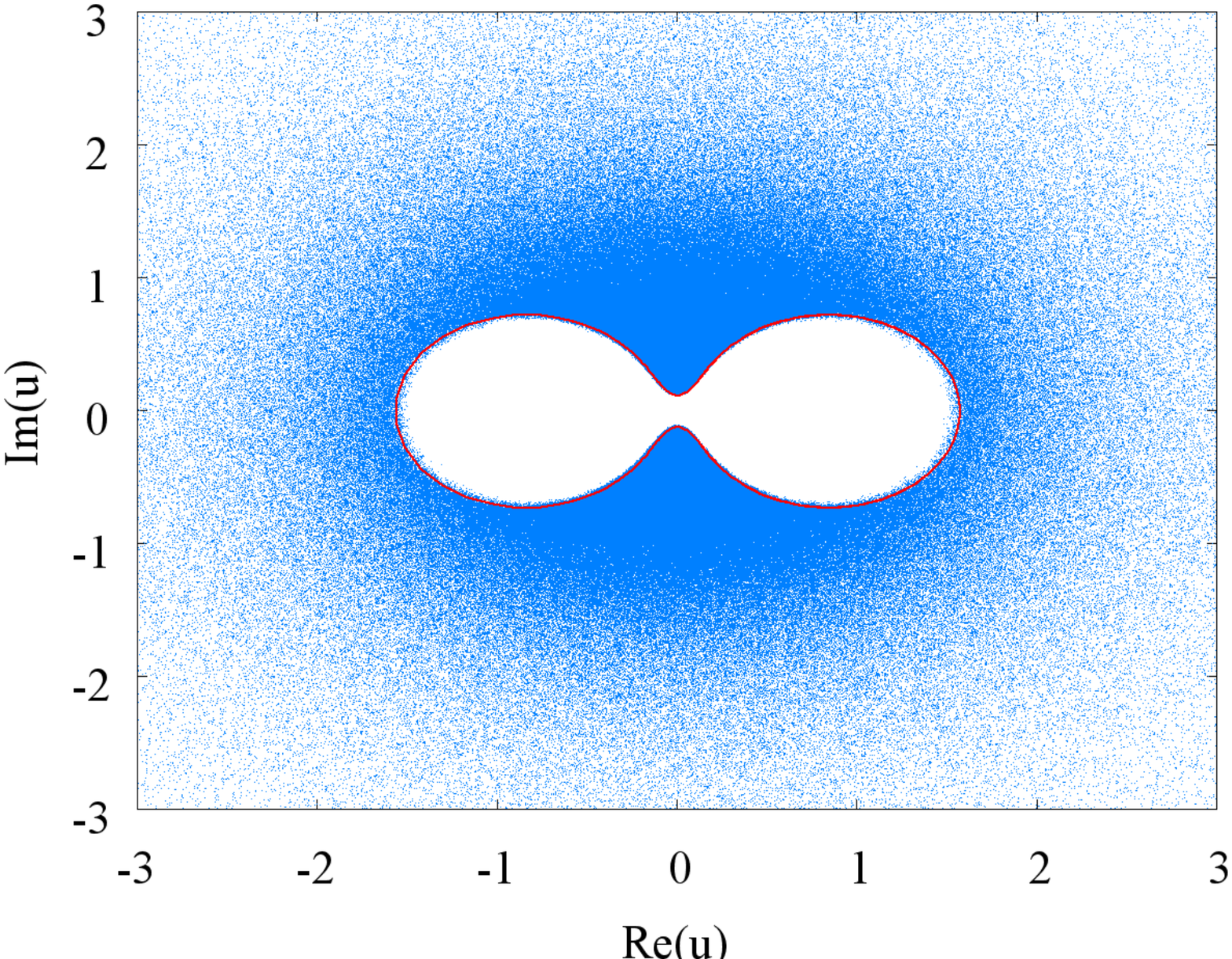} 
\\
\includegraphics[height=0.49\textwidth,angle=90]{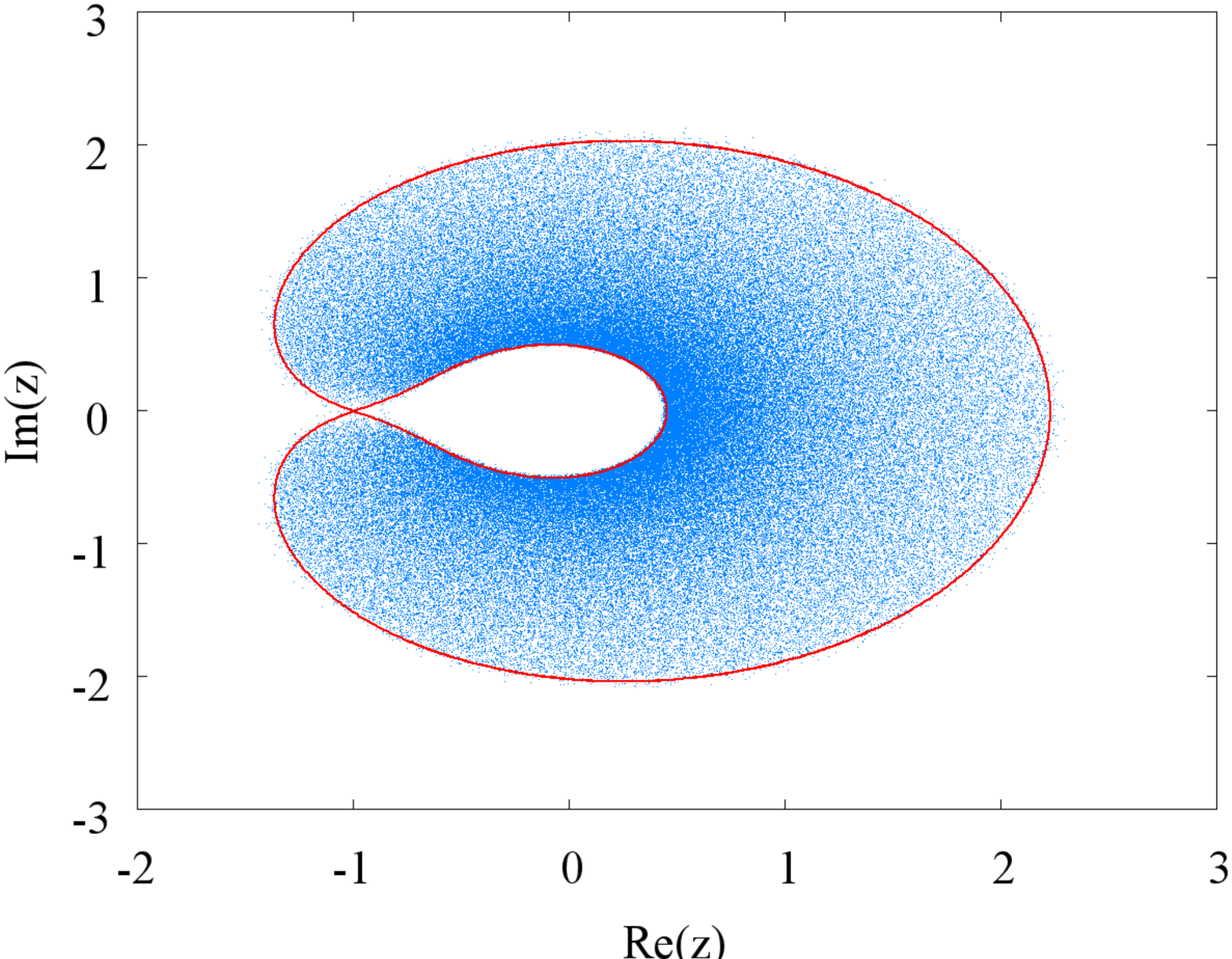}\hfill    \includegraphics[height=0.49\textwidth,angle=90]{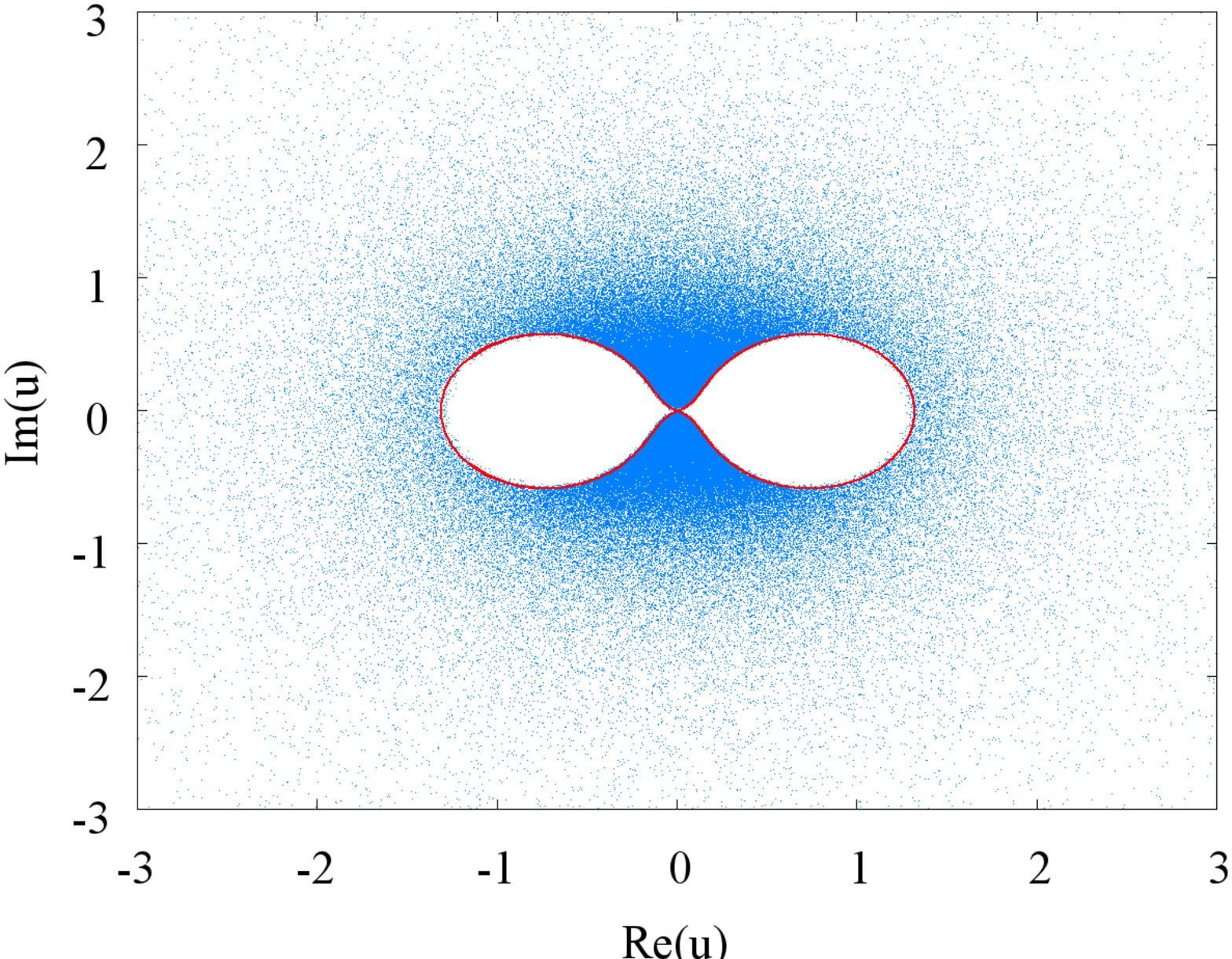}
\\
\includegraphics[height=0.49\textwidth,angle=90]{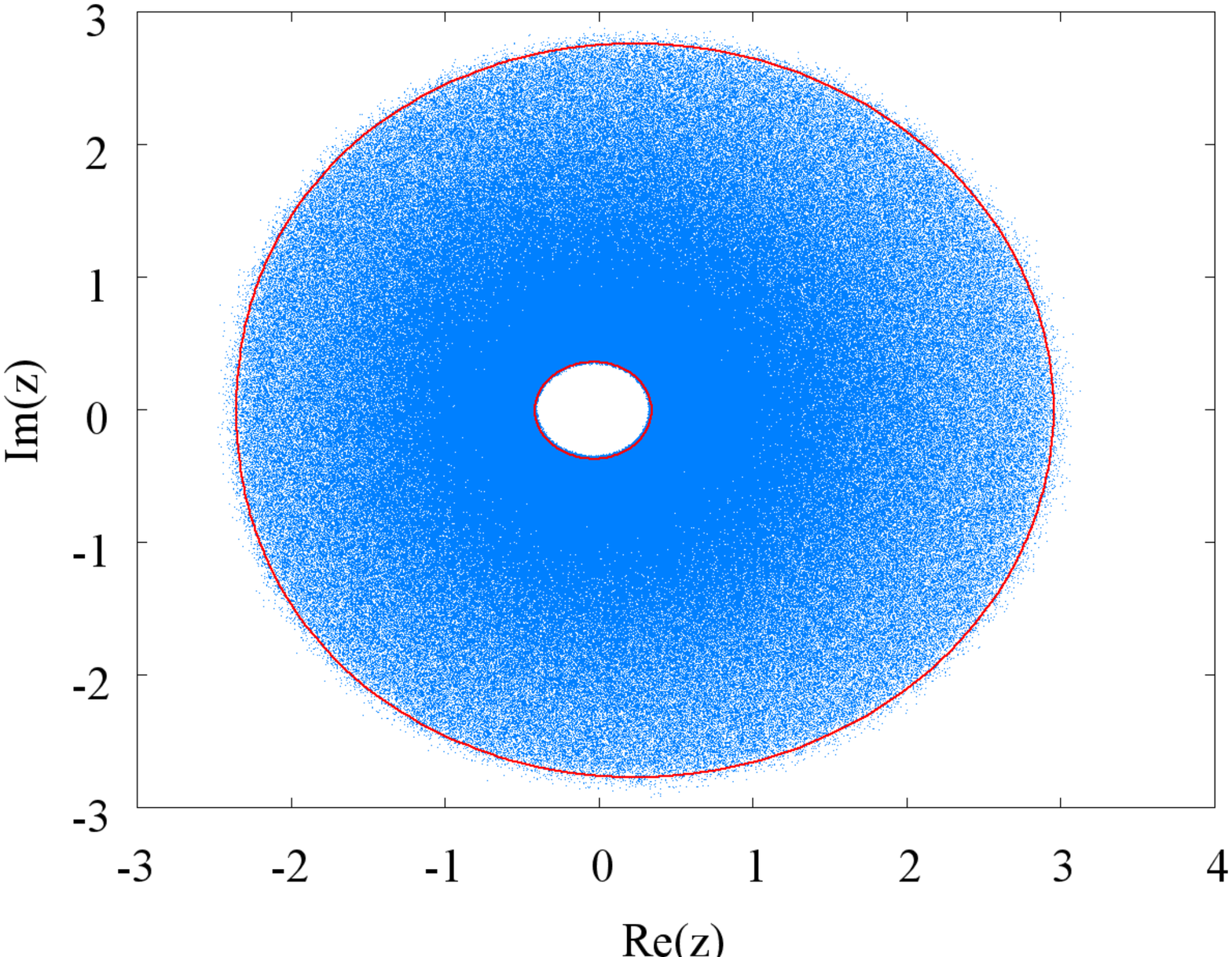}\hfill    \includegraphics[height=0.49\textwidth,angle=90]{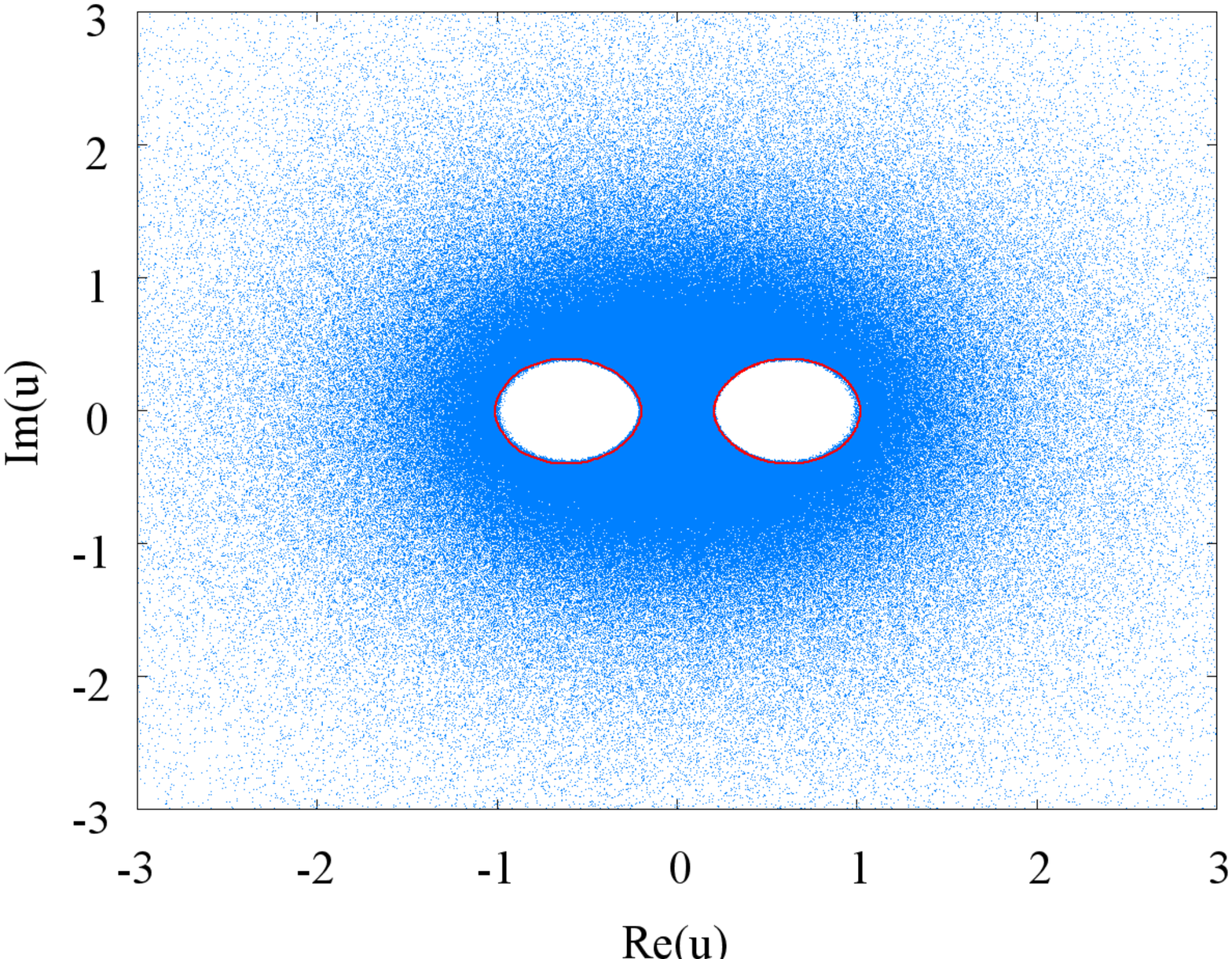} 
   \caption{Scatter plot of the eigenvalues of $W$ in the $z$-plane (left) and in the $u$-plane (right) for $\omega_1=1/{10}$, $\omega_2={1}/{2}$ and $t=5$ (top), $t=20/3$ (center), $t=10$ (bottom).}
\label{figS1}
\end{figure}

Figure~\ref{figS1} shows results of numerical
simulations for $\omega_1=1/10$, $\omega_2=1/2$, with all other
parameters as in Sec.~\ref{sec_NumRes}. The topological transition occurs at $t=20/3$, which corresponds to $t_+=\frac{20}{3}\left(\frac12+\frac1{10}\right)=4$ in agreement with the prediction.

Figure \ref{figS2} is generated with $\omega_2=1$ and $\omega_1=1/2000$ for $t=1$. Since this is already close to the unitary model, the eigenvalues are restricted to the vicinity of the unit circle in the $z$-plane, which corresponds to the imaginary axis in the $u$-plane. As $t_+$ is below the critical value we get no charge around $z=-1$ or $u=0$.

\FIGURE[ht]{
\centering
\includegraphics[height=0.49\textwidth,angle=90]{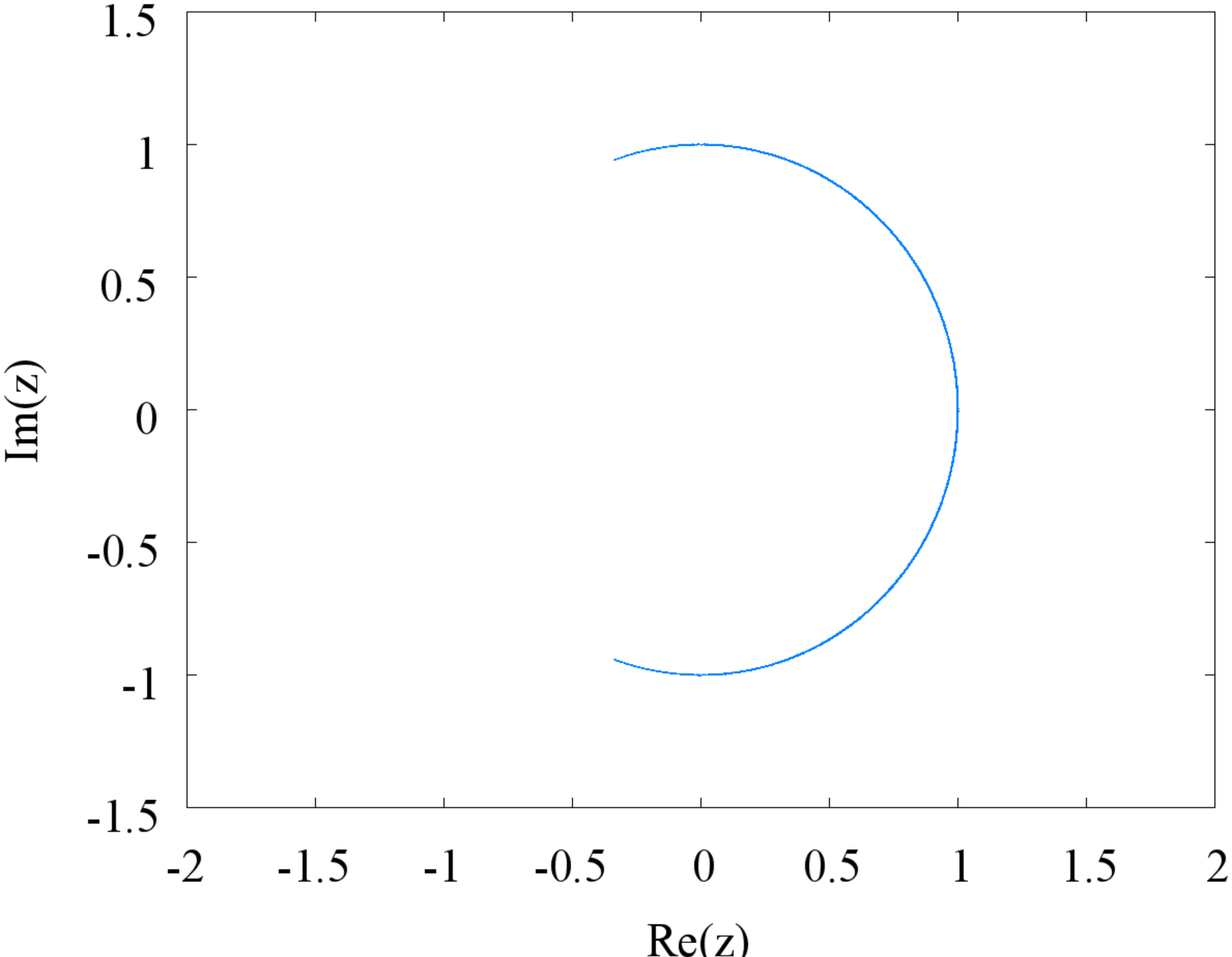}\hfill    \includegraphics[height=0.49\textwidth,angle=90]{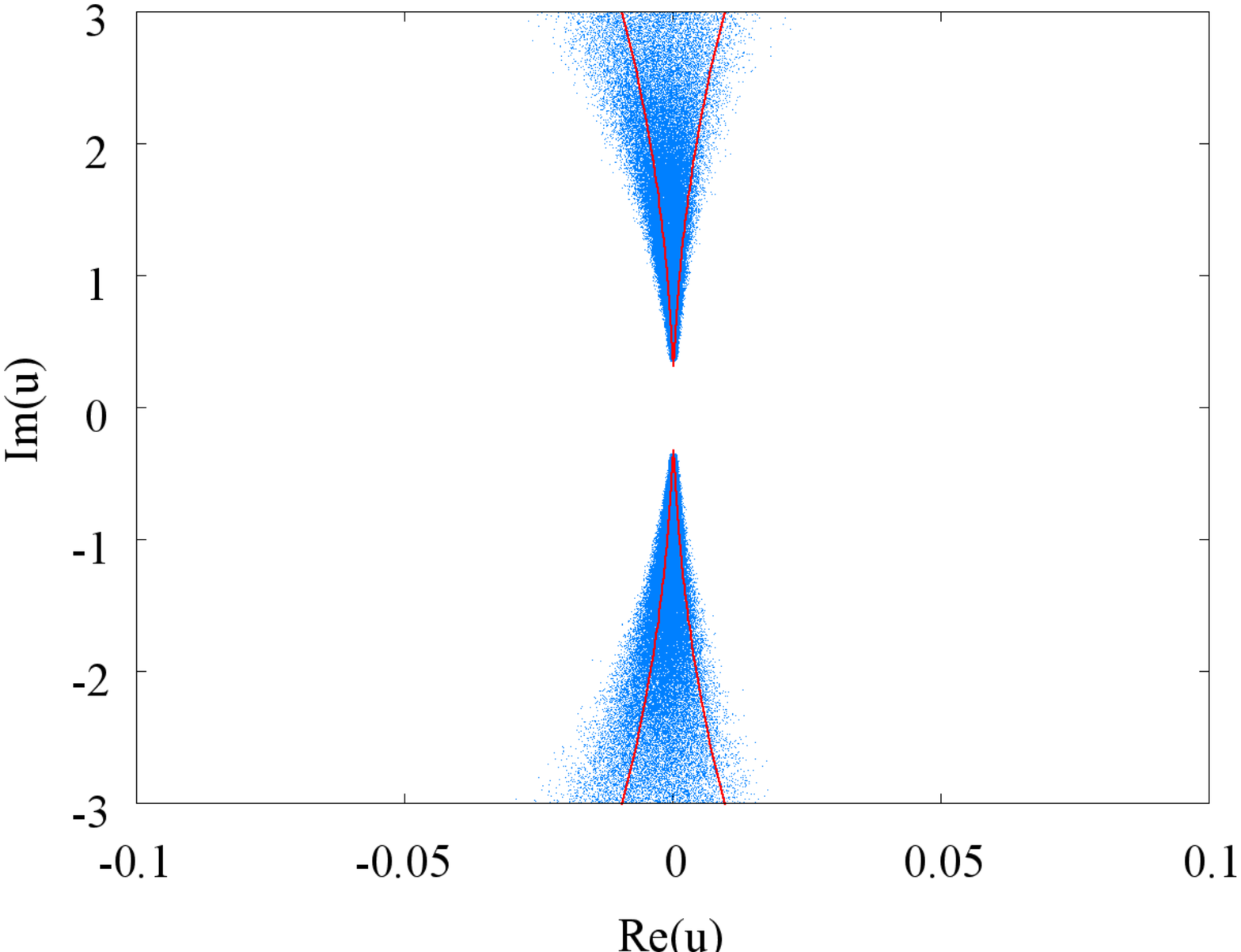}
\caption{Scatter plot of the eigenvalues of $W$ in the $z$-plane (left) and in the $u$-plane (right) for $\omega_1=1/{2000}$, $\omega_2=1$ and $t=1$.}  
\label{figS2}
}

\section[Beyond infinite $N$ and the associated saddle-point
approximation]{\boldmath Beyond infinite $N$ and the associated
  saddle-point approximation}

As mentioned in the introduction, we re-analyzed the model of~\cite{cmplxrmt} because we guess that it is a universal
representative of the large-$N$ phase structure of large 
classes of complex matrix Wilson loops. We shall refer to this
hypothetical class as the large-$N$ universality class. 

Like in the unitary case, we would like to study in more detail the approach to infinite $N$
and see what the matrix model universal features of this transition are. To do this we need a more
convenient finite-$N$ representation of the average of the product of two characteristic polynomials we have been looking at. More precisely, we
would like to first take the continuum limit $\epsilon\to 0$ without making any 
assumptions about how large $N$ is, and only later take $N$ large. 
We turn to an outline of how one could
achieve this; however we have not carried this to completion because, as
will become clear, a full analysis keeping full exact $N$-dependence is
complicated. We hope that using our technique 
one could learn which subleading terms in
the large-$N$ limit can be dropped without changing the universal 
properties of the approach to the large-$N$ limit. We start with the unitary case where the problem has been solved in~\cite{three-d} and present an alternative way
of deriving that solution. This alternative way has the potential to generalize to
the complex matrix case. We first look at the simplest complex matrix model and
then at the more general one. 

\subsection{The unitary case}

As before we consider the $SU(N)$ case but drop some
irrelevant $O(1/N)$ corrections to keep the formulas simple. 
When considering products of unitary matrices it suffices to look
at the average of the characteristic polynomial, and there is no need to
calculate the average of its absolute value squared. This is a 
significant simplification. 
Employing a quark 
representation (for details see~\cite{three-d}) we found the following representation of the average
characteristic polynomial,
\be
\langle \det ( z -W )\rangle =\int \prod_{i=1}^n \left [
\frac{\sqrt{N}d\lambda}{\sqrt{2\pi}} \right ] \left ( z-\prod_{j=1}^n
(1-\frac{\epsilon^2}{2} -\epsilon \lambda_j ) \right )^N
e^{-\frac{N}{2} \sum_{i=1}^n \lambda_i^2}\:.
\ee
Here, $W$ is the product of $n$ unitary matrices all close to the unit matrix. 
Introduce the notation
\be
\rho(\lambda)d\lambda = \sqrt{\frac{N}{2\pi}}
e^{-\frac{N}{2}\lambda^2} d\lambda
\ee
and
\be
a_n=\prod_{j=1}^n
(1-\frac{\epsilon^2}{2} -\epsilon \lambda_j )\:.
\ee
We are only interested in the limit $n\to\infty$, $\epsilon\to 0$ with
$t=n\epsilon^2$ kept fixed. We eliminate the $\lambda$-independent
term without affecting the limit by introducing new variables
\be
\hat a_n = \prod_{j=1}^n
(1-\epsilon \lambda_j ),~~~~~a_n=e^{-\frac{t}{2}} \hat a_n\:.
\ee

We now proceed by finding the probability density distribution for
the variable $\hat a_n$. In other words, we look for a way to perform
the integral over all $\lambda$-variables keeping the product we are
interested in fixed at some arbitrary value $\hat a$. This can be done in the
limit we are interested in. There, the probability density for $\hat a$
would be $P(\hat a;t)d\hat a$. $P(\hat a;t)d\hat a$ 
is obtained from the $P_n (\hat a) d\hat a$, the
probability densities  governing the variables $\hat a_n$ at step $n$.  
The basic step is to derive a recursion relation for the $P_n (\hat a)$ and
take the limit on that recursion relation. In the limit the recursion
relation becomes a partial differential equation of first order in $t$
and second order in $\hat a$. The reason for being first order in $t$ is that 
$P_n (\hat a )$ is determined by $P_{n-1} (\hat a )$; it does not
depend on $P_k (\hat a )$ with $k< n-1$. This reflects a ``Markov property''
of the multiplicative structure of the original matrix ensemble, and hence is 
a fundamental property~\cite{beni}. 
The partial differential equation is of Fokker-Planck type.  
In conjunction with the
boundary condition $\lim_{t\to 0^+} P(\hat a;t)= \delta(\hat a-1)$ the equation 
determines $P(\hat a;t)$ completely.  

Suppose we are at step $n$,
\be
\hat a_n = \hat a_{n-1} ( 1-\epsilon \lambda_n )\:.
\ee
We now drop the subscript $n$, i.e., $\hat a_n = \hat a$, $\lambda_n=\lambda$ and
$\hat a_{n-1} = \hat a+\delta \hat a$. To the order to which we need to keep terms we 
have
\be
\delta \hat a = -\hat a+\frac{\hat a}{ 1-\epsilon \lambda}= \hat 
a(\epsilon\lambda+\epsilon^2\lambda^2)\:.
\ee
Probability logic tells us that the probability to find $\hat a_n$ between $\hat a-\frac{\Delta \hat a}{2}$ and $\hat a+\frac{\Delta \hat a}{2} $ is given by 
\be
\prob \Bigl( \hat a-\frac{\Delta \hat a}{2} < \hat a_n <  \hat a+\frac{\Delta \hat 
a}{2}\Bigr) = \int d\lambda \rho (\lambda) \prob \Bigl( \hat a-\frac{\Delta \hat a}{2} < 
\hat a_{n-1}(1-\epsilon\lambda) <  \hat a+\frac{\Delta \hat a}{2}\Bigr)\:.
\ee
Probability densities depend on the measure,
\be
\prob \Bigl( \hat a-\frac{\Delta \hat a}{2} < \hat a_n <  \hat a+\frac{\Delta \hat 
a}{2}\Bigr) = P_n (\hat a)\Delta \hat a\:.
\ee
We end up with the recursion
\be
P_n (\hat a) =\int d\lambda \rho(\lambda ) P_{n-1} (\hat a+\delta \hat a)
(1+\epsilon\lambda+\epsilon^2\lambda^2)\:,
\ee
where the last factor comes from the change in measure.  
We now expand to second order in $\epsilon$ and do the $\lambda$-integral,
\be
P_n(\hat a)=P_{n-1}(\hat a) +\frac{\epsilon^2}{N} \left [ P_{n-1} (\hat a)
  +2\hat a\frac{\partial P_{n-1}(\hat a)}{\partial \hat a} +\frac{1}{2} \hat a^2
    \frac{\partial^2 P_{n-1}(\hat a)}{\partial \hat a^2} \right ].
\ee
We finally can take the limit,
\be
N\frac{\partial P(\hat a;t)}{\partial t}  = P(\hat a;t) + 2 \hat a
\frac{\partial P(\hat a;t)}{\partial \hat a} + \frac{1}{2} \hat a^2 
\frac{\partial^2  P(\hat a;t)}{\partial \hat a^2}\:.
\label{limitP}
\ee
This equation can also be written in the form
\be
\label{fpex}
N\partial_t P = \partial_{\hat a} (\hat a P) + \frac{1}{2} \partial_{\hat a}\hat 
a^2\partial_{\hat a} P\:,
\ee
showing explicitly that the integral $\int P d\hat a$ is time-independent
and therefore equal to unity, its initial value.  

The equation with the delta function initial condition has the
following solution~\cite{beni}, describing a log-normal distribution, 
\be
P(\hat a;t)=\sqrt{\frac{N}{2\pi t}} e^{-(\log \hat a +\frac{t}{2N})^2
  \frac{N}{2t} } \:.
\ee
We can now write an expression for the average characteristic
polynomial,
\be
\langle \det ( z -W )\rangle =\int d\hat a \left ( z- \hat a e^{-\frac{t}{2}}
  \right )^N P(\hat a;t)\:.
\ee
We change integration variables,
\be
\hat a = e^{-t\mu -\frac{t}{2N}}\:, 
\ee
to finally obtain
\be
\langle \det ( z -W )\rangle =\sqrt{\frac{Nt}{2\pi}} \int d\mu 
\left ( z-e^{-\frac{t}{2}-t\mu-\frac{t}{2N}}
 \right )^N e^{-\frac{Nt\mu^2}{2}}\:.
\ee
After dropping the $t/2N$ term in the exponent inside the parenthesis in
the integrand we obtain the result of~(\ref{jza}).  

One does not need to solve 
the Fokker-Planck equation exactly in order 
to get the large-$N$ limit, because 
$N$ plays the role of $1/\hbar$ (with Euclidean time though)
in~(\ref{fpex}). At large $N$ the Fokker-Planck equation
reduces to a Hamilton-Jacobi equation for $S$ with $P=e^S$,  
\be
N\frac{\partial S}{\partial t} = \frac{1}{2} \hat a^2 \left (
  \frac{\partial S}{\partial \hat a} \right )^2\:.
  \label{HamJac}
\ee
The solution that satisfies the initial
condition is particularly simple,
\be
S=-\frac{N}{2t} \log^2 \hat a \:.
\ee
As usual, there is a prefactor to $e^S$ which contains additional
$t$-dependence, but this factor is not needed for the large-$N$ limit.   
There are a few other terms which can
be ignored in the large-$N$ limit. 

In the expression for the average characteristic polynomial we now
have two terms that are exponential in $N$,
\be
e^{N\left [ \log (z-e^{-\frac{t}{2}}{\hat a}) -\frac{1}{2t}\log^2 \hat a\right 
]}\:.
\ee
The saddle-point
equation is
\be
-\frac{e^{-\frac{t}{2}}}{z-e^{-\frac{t}{2}}\hat a}=\frac{1}{t}\frac{\log
  \hat a}{\hat a}\:.
\ee
Define now $\hat a=e^{-t\lambda}$ and rearrange the saddle-point equation
slightly to obtain
\be
\lambda=-\frac{1}{1-ze^{t\lambda +\frac{t}{2}}}\:.
\ee
This reproduces the saddle-point equations we had before. 

The main conclusion is that, as it often is the case in the context of
large-$N$ models, one has a
``quantum''-like equation for finite $N$, with $1/N$ playing a
role analogous to $\hbar$. The large-$N$ limit is then
``semiclassical'', with the ``quantum'' equation being replaced by a
classical one, in a variant of the WKB method.  

This is what we would like to duplicate in the complex matrix case.

\subsection{The basic product of random complex matrices}

\subsubsection{An exact map to a product of random $2\times 2$ matrices}

In the following, we do not restrict the trace of $W_n$ to keep the analysis as simple as possible.
As explained in Sec.~\ref{sec_saddleptanalysis}, we can then start from Eq.~(\ref{keya}) without the $\lambda$-integrals,
\begin{align}
\la|\det(z-W_n)|^2\ra&=\NAn\int\prod_{j=1}^n [d\bar\psi_j d\psi_j d\bar\chi_j
d\chi_j d\mu(\zeta_j)]   e^{-\epsilon\sum_{j=1}^n (\zeta_j\bar\psi_j \chi_{j-1} - \zeta^*_j
\bar\chi_{j}\psi_{j+1})} \nonumber\\
&\quad\times
 e^{-\sum_{j=1}^n( \bar\psi_j \psi_{j+1} 
+\bar\chi_j \chi_{j-1}) }  
e^{\sum_{j=1}^n (e^\sigma \bar\psi_j\psi_{j} + e^{\sigma^*}
  \bar\chi_j \chi_{j}) }e^{-N\sum_{j=1}^n|\zeta_j|^2}\:.
  \label{keyac}
\end{align}
We now change notation by introducing two-component 
Grassmann variables $\Phi_j$, $\bar\Phi_j$,
\be
\Phi_j=\begin{pmatrix}\psi_j\\\chi_{j-1}\end{pmatrix},~~~~\bar\Phi_j=\begin{pmatrix}\bar\psi_j & \bar\chi_j\end{pmatrix}\:.
\ee
Equation~(\ref{keyac}) can then be written as 
\begin{align}
\la|\det(z-W_n)|^2\ra&=(-1)^{N(n-1)} \int\prod_{j=1}^n [d\bar\Phi_j d\Phi_j d\mu(\zeta_j)] e^{-N\sum_{j=1}^n|\zeta_j|^2}\nonumber \\
&\quad\times 
\exp\Biggl[ \sum_{j=1}^n\left ( \bar\Phi_j\begin{pmatrix}e^\sigma & -\epsilon \zeta_j\\
0 & -1\end{pmatrix} \Phi_{j} 
+\bar\Phi_j \begin{pmatrix}-1&0\\ \epsilon \zeta_j^* & e^{\sigma^*} \end{pmatrix} \Phi_{j+1}\right ) \Biggr ], \label{keyb}
\end{align}
where the factor $(-1)^{N(n-1)}$ results from bringing the integration measures in canonical order after the index shift in $\chi_j$.

We can again change integration variables from $\bar\Phi_j$ 
to  $\bar\Phi_j\begin{pmatrix}e^\sigma & -\epsilon \zeta_j\\ 0 & -1\end{pmatrix}$. Now, 
using the identity~(\ref{quark}), we can integrate over all the Grassmann variables, obtaining
an expression containing 
another random matrix product, but this time the matrices are just $2\times 2$.
\be
\la|\det(z-W_n)|^2\ra=(-z)^N \int\prod_{j=1}^n [d\mu(\zeta_j)] e^{-N\sum_{j=1}^n|\zeta_j|^2}
{\det}^N\left ( 1 - \frac{1}{z}\prod_{j=1}^n Q_j\right ),
\label{keybQ}
\ee
where
\be
Q_j=\begin{pmatrix} 1+\epsilon^2 |\zeta_j|^2& \epsilon e^{\sigma^*}\zeta_j\\ \epsilon\zeta_j^* 
e^\sigma & e^{\sigma+\sigma^*}\end{pmatrix}.
\ee
Absorbing some phase factors into the variables $\zeta_j$, the matrices $Q_j$ can be further simplified to
\be
Q_j=
e^{\frac{\sigma+\sigma^*}{2}}\begin{pmatrix} (1+\epsilon^2 |\zeta_j|^2)e^{-\frac{\sigma+\sigma^*}{2}}& \epsilon \zeta_j\\ \epsilon\zeta_j^* & e^{\frac{\sigma+\sigma^*}{2}}\end{pmatrix}.
\ee
The matrix after the prefactor is an $SL(2,\mathbb{C})$ matrix. Hence, the product of $Q_j$ matrices will
be, up to a multiplicative factor, also in $SL(2,\mathbb{C})$. The new multiplicative random
matrix model defines a stochastic process on the $SL(2,\mathbb{C})$ manifold. Unlike the 
$N\times N$ matrix ensemble, the $2\times2$ matrix ensemble has no inversion symmetry and
only a restricted conjugation symmetry. 

Finally, we write
\be
\la|\det(z-W_n)|^2\ra=(-z)^N \int\prod_{j=1}^n [d\mu(\zeta_j)] e^{-N\sum_{j=1}^n|\zeta_j|^2}
{\det}^N\left ( 1 - e^{-i\Psi}\Delta_n\right )\:.
\ee
Here, $\Delta_n=\prod_{j=1}^n Y_j$ and
\be
Y_j= \begin{pmatrix}|z|^{-1/n} (1 +\epsilon^2 |\zeta_j|^2) & \epsilon\zeta_j\\ \epsilon\zeta_j^* &
|z|^{1/n}\end{pmatrix}.
\ee
The new multiplicative matrix ensemble has $\det Y_j =1$, so $\Delta_n$ is 
restricted to $SL(2,\mathbb{C})$, which in turn implies
\be
\la|\det(z-W_n)|^2\ra=|z|^N \int\prod_{j=1}^n [d\mu(\zeta_j)] e^{-N\sum_{j=1}^n|\zeta_j|^2}
\left ( \tr \Delta_n - 2\cos\Psi \right )^N\:.
\ee
In the new ensemble we have invariance under complex conjugation and conjugation by 
a $U(1)$ subgroup,
\be
Y_j \rightarrow \begin{pmatrix}e^{i\theta_j} & 0 \\ 0 & e^{-i\theta_j}\end{pmatrix} Y_j 
\begin{pmatrix}e^{-i\theta_j} & 0 \\ 0 & e^{i\theta_j}\end{pmatrix}.
\ee
Therefore, the probability density of $\Delta_n$ will be invariant under
\be
\Delta_n \rightarrow \begin{pmatrix}e^{i\theta_n} & 0 \\ 0 & e^{-i\theta_n}\end{pmatrix} \Delta_n  \begin{pmatrix}e^{-i\theta_n} & 0 \\ 0 & e^{i\theta_n}\end{pmatrix}.
\ee
So, the FP equation for this case,
\be
\frac{\partial \Sigma_N}{\partial t} = \Theta_N \Sigma_N\:,
\ee
is an equation for a function $\Sigma_N$ of six real variables including $t$ ($z$ is a parameter). $\Theta_N$ is a linear partial differential operator of second degree in five real variables. There are no terms
from the measure if we pick the latter to be $SL(2,\mathbb{C})$ invariant. 

The dependence on $N$ is explicit. All second-order derivative terms
carry a $1/N$ factor. Among the first-order derivative terms some have a
$1/N$ factor and others are $N$-independent.

One can write down an exact integral expression for $Q(z,z^*)$ in terms of
$\Sigma_N (\Delta_n;|z|,t)$,
\be
Q(z,z^*;t)=|z|^N \int d\mu(\Delta_n ) \Sigma_N (\Delta_n ;|z|;t) (\tr \Delta_n -2 \cos\Psi)^N\:.
\label{keyc}
\ee
In the infinite-$N$ limit, $Q(z,z^*)$ would be given by a
dominating saddle point, and subleading corrections would identify the relevant large-$N$ universality class. 

An analysis of the Fokker-Planck equation for arbitrary $z$ seems too complicated to
attack directly, so we restrict our attention to the unit circle, $|z|=1$.

\subsubsection[The case of $|z|=1$: Simplifications for large
$N$]{\boldmath The case of $|z|=1$: Simplifications for large $N$}

We now set $|z|=1$. This simplifies the operator $\Theta_N$, eliminating the first-order
derivative terms that had no $N$-dependence. Consequentially, at $|z|=1$ one has
\be
N\frac{\partial\Sigma_N}{\partial t} = \Theta \Sigma_N\:,
\ee
where $\Theta$ has no $N$-dependence.

We are looking for solutions having the structure
\be
\Sigma_N\sim\exp{\left[-\frac{N}{t}S +\ldots\right]},
\label{wkb}
\ee
where the dots stand for terms subleading in $t/N$. Looking at the structure of
the FP equation, we see that all terms linear in the derivatives can be ignored at large $N$
as long as we keep $|z|=1$. 
This leads us to replace the $|z|=1$ model, defined by the factors
\be
Y_j= \begin{pmatrix} (1 +\epsilon^2 |\zeta_j|^2) & \epsilon\zeta_j\\ \epsilon\zeta_j^* & 1\end{pmatrix},
\ee
by a new model, defined by the factors
\be
Y^\prime_j= \frac{1}{\sqrt{1-\epsilon^2 |\zeta_j|^2}}\begin{pmatrix} 1 & \epsilon\zeta_j\\ \epsilon\zeta_j^* &1\end{pmatrix}.
\ee
The $Y^\prime$ model preserves the $U(1)$ symmetry of
the $Y$ model and differs from it only in the first-order derivative terms, while the powers
of $N$ appear in the same places as before. Therefore, the leading large-$N$ behavior of
the two models is the same.  
The advantage of the new model is that the $Y^\prime_j$ are restricted 
to an $SU(1,1)$ subgroup of $SL(2,\mathbb{C})$, which forces also the 
product $\Delta^\prime$ into $SU(1,1)$. 
This implies that the  
solution $\Sigma^\prime_N (\Delta^\prime; t)$ depends, in addition to $t$, on only 
two real variables. 

We observe now that with the right choice of variables
the second-order derivatives acting on $\Sigma_N(\Delta; t)$ only attack two of the five
real arguments (on which $\Sigma_N$ depends, in addition to $t$) also in the $Y$ model. 
Therefore, in the large-$N$ limit,
one can again look for a solution of the form~(\ref{wkb}). Further, this discussion indicates that
significant simplifications will occur at large $N$ even for $|z|\ne 1$, when first-order
derivative terms that matter also in the large-$N$ limit appear.

\subsubsection[The case of $|z|=1$: The Fokker-Planck (FP)
equation]{\boldmath The case of $|z|=1$: The Fokker-Planck equation}

We now focus on the $SU(1,1)$ model and for notational convenience drop the primes, which
now get a different use. 
The recursion relation determining the FP equation is
\be
\Delta=\Delta^\prime Y
\ee
with 
\be
Y= \frac{1}{\sqrt{1-|\omega|^2}}\begin{pmatrix} 1 & \omega\\ \omega* &1\end{pmatrix},~~~\omega=\epsilon\zeta\:.
\ee

The structure of the FP equation will be simpler in a well-chosen
parametrization. The best choice of parameters 
is determined by the symmetries
obeyed both by the equation and by our particular initial condition.

The recursion relation has an obvious invariance $\Delta\to S\Delta$, $\Delta^\prime\to S\Delta^\prime$
with $S\in SU(1,1)$. This means that the FP equation for $\Delta$ will be invariant under
multiplication from the left. In addition,
because of the symmetry of the distribution of $Y$ under conjugation by 
the diagonal $U(1)$ subgroup of $SU(1,1)$, there is an invariance of the equation under
right multiplication by these group elements. Both of these invariances are broken by the initial condition and therefore parameters that would
be associated with these invariances cannot be eliminated. 

However, the diagonal $U(1)$ subgroup of the left $SU(1,1)$
combined with the right $U(1)$ produces an $U(1)\times U(1)$
subgroup, and the initial condition is invariant under its diagonal subgroup.
Thus, there is only one $U(1)$ under which both the equation and the initial condition are invariant.
This allows us to eliminate one out of the three variables parametrizing $SU(1,1)$, making  
the operator $\Theta$ on the right-hand side of the FP equation a second-order partial differential
operator in two real variables. However, the form of $\Theta$ is much more restricted, 
because $\Theta$ does not know about the initial condition, and obeys many more symmetries. 
$\Theta$ must descend from the invariant Laplacian on the $SU(1,1)$ group manifold after
the elimination of the one extra variable. The structure of the Laplacian must be such that this
restriction is compatible with the partial differential equation based on $\Theta$. 
We do not get evolution on a coset $SU(1,1)/U(1)$
because the variable we can eliminate is determined by the initial boundary condition, 
which singles out a specific $U(1)$ subgroup that couples the right and left invariances. 

A direct derivation of the FP equation confirms the considerations
above.
We parametrize $\Delta$ by
\be
\Delta=\begin{pmatrix}a & b \\ b^* & a^* \end{pmatrix},~~~|a|^2-|b|^2=1\:,
\ee
with
\be
a=\sqrt{u} e^{i\phi},~~b=\sqrt{u-1} e^{i\psi},~~~\infty\ge u \ge 1,~-\pi\le \phi,\psi\le\pi\:.
\ee
In these variables the invariant measure on $SU(1,1)$ is $du d\phi d\psi$, up to a constant. 
The recursion relation for $\Delta$ is 
\be
\begin{pmatrix}a-\delta a & b-\delta b\\ b^*-\delta b^* & a^*-\delta a^* \end{pmatrix}=\begin{pmatrix}a&b\\ b^*& a^*\end{pmatrix} Y^{-1}\:,
\ee
where $\delta a = a - a^\prime$ and $\delta b = b- b^\prime$.
Working out the algebra, and keeping only terms up to second order in $\omega$ and among those
only terms that could contribute to a term of the form $|\omega|^2$, we get
\be
\delta u = -(2u-1) |\omega|^2 +\sqrt{u(u-1)} \left ( e^{i(\psi-\phi)} \omega^* + e^{-i(\psi-\phi)} \omega\right )
\ee
and
\be
\delta\phi=\frac{i}{2}\sqrt{\frac{u-1}{u}}\left ( e^{i(\phi-\psi)} \omega - e^{i(\psi-\phi)} \omega^* \right ).
\label{deltaPhi}
\ee
The FP equation is then
\be
N\frac{\partial \Sigma_N (u,\phi;t) }{\partial t} = - H \Sigma_N (u,\phi;t)\:,
\label{fpa}
\ee
where $H$ is a non-negative hermitian operator,
\be
H=-\frac{\partial}{\partial u} u(u-1) \frac{\partial}{\partial u} - \frac{u-1}{4u}\frac{\partial^2}
{\partial \phi^2}\:.
\ee
Transforming variables to $u=(1+x)/2$, $x\ge 1$, we get
\be
H=-\frac{\partial}{\partial x} (x^2-1) \frac{\partial}{\partial x} - \left ( \frac{x}{2(x+1)}-\frac{1}{4}\right ) \frac{\partial^2}
{\partial \phi^2}\:.
\label{Hxphi}
\ee
This equation is almost identical to Eq.~(28) in~\cite{beni}. The difference is the
$1/4$ term in the prefactor of the second derivative with respect to $\phi$. The invariances of the equation are the reason for the similarity. In~\cite{beni} the 
multiplicative random ensemble consists of real $2 \times 2$ matrices of the form 
$1+\epsilon X$, where $X$ is real and drawn from identical Gaussian distributions for each of 
its four entries. By factoring the determinant this would provide an evolution on 
the group $SL(2,\mathbb{R})$ or, more precisely, $SO(2,1)$. $SL(2,\mathbb{R})$ and $SU(1,1)$ are the same
for our continuous evolutions emanating from the identity matrix. The
left invariances are the same in both cases since they are 
independent of the distribution of the individual factors.
The difference we have obtained must therefore reflect 
the difference in the right invariances, which do depend on the
distributions of the individual factors. Thus, one gets slightly different restrictions of the
heat-kernel equation. 

The initial condition $\lim_{t \to 0^+} \Sigma_N(u,\phi,t) =2 \delta(u-1)\delta_{2\pi} (\phi)$
 ($\delta_{2\pi}(\phi)=\frac{1}{2\pi}\sum_{n\in \mathbb{Z}} e^{in\phi}$) 
 is $N$-independent, and we take the integration measure to be $du d\phi$. Taking into account Eq.~(\ref{fpa}), we conclude that the dependence of $\Sigma_N$ on $N$ and $t$ is of the form
\be
\Sigma_N(u,\phi;t)=\Sigma\left(u,\phi;\frac{t}{N}\right).
\ee
Therefore, the large-$N$ limit is dominated by the short time ($\hat t=t/N$) behavior of the 
probability distribution $\Sigma(u,\phi;\hat t)$. 

In~\cite{beni} the authors solve their equation by separation of variables. The 
$\phi$-dependence must be periodic, and is labeled by an integer $m\in \mathbb{Z}$. In each sector,
$H$ is replaced by
\be
H_m=H=-\frac{\partial}{\partial x} (x^2-1) \frac{\partial}{\partial x} + m^2 \left ( \frac{x}{2(x+1)}-\frac{1}{4}\right ) .
\ee
The eigenfunctions and eigenvalues of $H_m$ are known exactly. We see that our problem
will be solved in an identical way, only the eigenvalues have to 
be shifted by $(m/2)^2$. This shift does not affect the matching onto the initial condition, which is the same here as in~\cite{beni}. Therefore a
formula for $\Sigma(u,\phi;\hat t)$ is available, and we know that, although explicit, it is
difficult to do much with it at the analytic level. 


\subsubsection[The case of $|z|=1$: Large-$N$ limit from the
Fokker-Planck equation]{\boldmath The case of $|z|=1$: Large-$N$ limit
  from the Fokker-Planck equation}\label{sec_largeNfromFP}

For the modified model, Eq.~(\ref{keyc}) looks as follows for $z=e^{i\Psi}$,
\be
Q(z,z^*;t)=\int_1^\infty du \int_{-\pi}^\pi d\phi \Sigma\left(u,\phi;\frac{t}{N}\right) 2^N (\sqrt{u} - \cos\Psi )^N\:.
\ee
As $t/N \to 0$, $\Sigma$ must become a delta-function in $u$ and $\phi$. 
Therefore, for small $t/N$ we expect $\Sigma(u,\phi;t/N)$ to drop rapidly
as $u$ increases beyond 1 and $\phi$ departs from 0. Looking at $H$, we realize that
for $u$ close to 1 the $\phi$-derivative term is suppressed. This leads us to the simple
ansatz
\be
\Sigma\left(u,\phi;\frac{t}{N}\right)\sim \frac{N}{t} \delta_{2\pi} (\phi) e^{-\frac{N}{t} (u-1)}\:.
\ee
When this is inserted into the expression for $Q$ the integral over $\phi$ is trivial, leaving
only the integral over $u$. The latter will be dominated by a saddle point or by the endpoint
$u=1$. When the endpoint dominates we get the holomorphically factorized answer
$| 1 - e^{i\Psi}|^{2N}$ we have seen before. Thus, ``saddle A'' corresponds to  
endpoint dominance. The saddle-point equation for $u$ is
\be
\frac t2=\left(\sqrt{u}-\cos\psi\right)\sqrt{u}\:,
\ee
and its positive solution is given by
\be
\sqrt{u_s}=\frac12\left(\cos\psi+\sqrt{\cos^2\psi+2 t}\right).
\ee
We see that this saddle will first become available at a point $z=e^{i \Psi}$ on the unit circle only when
\be
\sqrt{u_s}>1,~~\cos \Psi>1-\frac t2\:,
\ee
in agreement with our findings earlier. 
Once the saddle is away from the endpoint the ansatz form of $\Sigma$
is no longer even
plausible, and a more complete analysis is needed.

\subsection{The generalized Gaussian model: Exact map to a random
  multiplicative model of $2\times 2$ matrices}


When $\omega_1\ne\omega_2$, that is in the generalized case, one can again
reduce the problem to a product of random $2\times 2$ matrices, albeit 
of a slightly more complicated structure than the one we have seen in the
$\omega_1=\omega_2$ case earlier. 

Using similar manipulations one derives the following representation,
\begin{align}
\langle |\det (z - W_n ) |^2 \rangle & = \NAn\NCn(-z)^N \int \prod_{j=1}^n [d\mu(\zeta_j )
d\xi_j d\theta_j e^{-N\sum_{j=1}^n (|\zeta_j|^2 +\frac{1}{2}\xi_j^2 +\frac{1}{2} \theta_j^2 )}\nonumber \\ 
&\quad\times \Biggl[ \prod_{j=1}^n (d_j) \Biggr ]^N
\Biggl [ \det \Biggl ( {\bf 1} -\prod_{j=1}^n ( A_j^{-1} B_j ) \Biggr ) \Biggr ]^N\:.
\end{align}
Here,
\be
d_j=1-\frac12 \omega_-^2-\omega_- \theta_j
\ee
and
\be
A_j = \begin{pmatrix}e^\sigma & \omega_+ \zeta_j\\ 0& 1-\frac12 \omega_-^2- \omega_- \theta_j\end{pmatrix},~~~~~
B_j=\begin{pmatrix}1-\frac12\omega_-^2-\omega_- \xi_j & 0\\ -\omega_+\zeta_j^* & e^{\sigma^*}\end{pmatrix}.
\ee
Now, one can proceed to take the $\epsilon\to 0$ limit, deriving a FP
equation for the new $2\times 2$ random matrix product of $A_j^{-1} B_j$. 
The structure is similar to the one in the special case analyzed before,
and no progress can be made before the special case is fully solved. 

\subsection[Large-$N$ universality]{\boldmath Large-$N$ universality}

The main objective of the attempt to go beyond the infinite-$N$
saddle-point approximation is to identify a universality class for the
large-$N$ 
phase transition, its exponents and its associated relevant perturbations.
For the case of $\langle\det(z-W)\rangle$ in the unitary case, this
has been achieved in previous work. For complex matrices we need the
more complicated object $\langle|\det(z-W)|^2\rangle$, which both has a 
large-$N$ phase transition and a region where large-$N$ factorization 
is not useful in the sense that  $\langle|\det(z-W)|^2\rangle \ne 
|\langle\det(z-W)\rangle|^2 $. We suspect that the large-$N$ universal
region will have to deal with both of these issues. A direct approach similar
to the unitary case seems difficult, but it is clear that one can make
simplifications that do not matter at large $N$ without loosing the universal
properties. We have not yet learned how to do this effectively and reliably. 

A simpler case might be when $\omega_1 t \ll 1$. In that case we are close to
the unitary case, with the unit circle slightly expanded into a strip
of similar shape in the complex plane. This case should be easier to treat,
in the sense of establishing large-$N$ universality in an appropriately
defined regime of ``weak non-unitarity''. This would be an analog of the regime
of weak non-hermiticity in non-multiplicative random complex matrix 
ensembles~\cite{fyod}.

\section{Summary}

The main objective of this paper was to generalize the universal large-$N$
phase transition occurring in field theory models, which have important 
non-local observables that can be thought of as fluctuating unitary matrices,
to the case where the non-local observables are more naturally chosen as 
complex matrices. In both cases a basic multiplicative structure is assumed,
where the observable is a matrix given by the product of many matrices 
close to unity. The individual factors are sufficiently decorrelated to
undergo a large-$N$ phase transition in the same large-$N$ universality class
as simple multiplicative matrix models, where the factors are completely uncorrelated. 
The phase transition occurs simply because the effective number of factors
grows beyond a certain limit when a regime of strong coupling is entered. 
The effective number of factors depends both on the true number of factors
and the departure from unity of each individual factor.

When the coupling is small, perturbation theory
indicates that parallel transport round a closed loop will be a matrix close to unity; when the coupling is large, the parallel transport is likely to depart from unity in a substantial manner. 
At infinite $N$ the $SL(N,\mathbb{C})$ random matrix
associated with parallel transport round a closed finite smooth curve will have a spectrum which
does not reach the origin of the complex plane, nor does it run away to infinity. Thus, we
can ask if the support of the spectrum separates zero from infinity or not. It is possible
that for small couplings it does not, but for large couplings it does. As the coupling is
varied from zero upwards a large-$N$ phase transition would then occur. Just like in the unitary
case, traces of finite powers of the Wilson loop matrix need not exhibit any nonanalyticity, but
the eigenvalue distribution of the Wilson loop matrix is singular at the transition.

A natural question is then what happens in a
conformal gauge theory, where there is no difference between small and large loops and where Wilson loops corresponding to complex matrices
naturally enter. It is quite possible that in very special cases~\cite{pestun} no
large-$N$ phase transition occurs. In such cases the behavior of the Wilson
loop is captured by single matrix models, and there is no hint of
non-commutativity  playing a role. 

The conformal case can have a dimensionless coupling constant which is scale-indepen\-dent and a free parameter of the theory. It is quite possible that less special Wilson loops do exhibit the above transition. 

Even in the unitary cases it may be advantageous to use a regularization that
introduces some weak non-unitarity, in which case a study of the universality
of the appropriate matrix model would be useful. 

While our work shows beyond reasonable doubt that the Durhuus-Olesen transition
of the unitary matrix case generalizes to complex matrices, more work is needed
to bring the understanding of large-$N$ universality in the complex case to the
same level as that of the unitary case. While the latter corresponds to 
a regularization~\cite{blaizot} of the singularity-generating mechanism of the inviscid Burgers equation for wave propagation restricted to 
some line, we now need something similar for
the inviscid Burgers equation for complex waves 
propagating in the complex plane. Much is known about this, and it is hard to
believe that the particular regularization employed by going to the large-$N$ 
``double-scaling'' limit will generate something totally new. To know what
exactly happens more work is needed, but the vast existing 
applied mathematics literature indicates that the problem is soluble. 
There exists an example in the context of single matrix models 
where a connection to the deformation theory
of some specific partial differential equations has been established~\cite{dubrovin}. 
We plan to return to this topic in the future.

\acknowledgments

We acknowledge support by BayEFG (RL), by the DOE under grant
number DE-FG02-01ER41165 at Rutgers University and by the SAS of Rutgers University (HN), and by DFG (TW).
HN notes with regret that his research has for a long time been 
deliberately obstructed by his high-energy colleagues at Rutgers.  
HN also acknowledges a Humboldt award which supported his stay 
at Humboldt University in Berlin where part of this research was carried out. 
Another part of this research was carried out during a visit at the Newton Institute in
Cambridge, UK. 
HN also gratefully acknowledges a conversation and some email exchanges with
R.A. Janik. Several useful conversations with J. Feinberg and R. Narayanan
are also acknowledged. 
In addition, HN expresses his thanks to M. {\" U}nsal who 
invited him to give a seminar at SLAC. Comments from
the audience encouraged his search for a complex generalization of the large-$N$ phase
transition away from the unitary matrix case. 

\bibliographystyle{jhep}

\bibliography{cmplx_v1}

\end{document}